\documentclass[aps,physrev,twocolumn,superscriptaddress]{revtex4-2}

\usepackage{xcolor}
\usepackage{graphicx}
\usepackage{hyperref}
\usepackage[caption=false]{subfig}
\usepackage{braket}
\usepackage{comment}
\usepackage{amsmath}
\usepackage{amssymb}
\usepackage{dsfont}

\begin{document}

\title{Towards Lattice Surgery Compilation\\for the Color Code Using Pipe Diagrams}

\author{Laura S. Herzog}
\email{laura.herzog@tum.de}
\affiliation{Chair for Design Automation, Technical University of Munich, Germany}

\author{Gilad Kishony}
\affiliation{Classiq Technologies, Tel Aviv, Israel}

\author{Robert Wille}
\affiliation{Chair for Design Automation, Technical University of Munich, Germany}
\affiliation{MQSC, Garching near Munich, Germany}

\author{Austin Fowler}
\noaffiliation

\begin{abstract}
Pipe diagrams have emerged as a powerful framework for flexible lattice surgery compilation and spacetime optimization for the surface code. In contrast, analogous compilation techniques for color code architectures remain largely unexplored, despite the color code’s favorable properties, including reduced qubit overhead and transversal single-qubit Clifford gates. In this work, we develop a pipe diagram representation for the triangular color code on the 6.6.6 lattice and establish its correspondence to ZX-diagrammatic descriptions of computation. We present distance-independent constructions of color code pipe diagrams together with explicit realizations of correlation surfaces, stabilizers, and syndrome extraction circuits. This framework enables both macroscopic optimization of logical computations in spacetime and microscopic compilation to executable syndrome extraction circuits. %
We demonstrate the potential for compact spacetime embeddings with the color code's geometry. %
These results provide a foundation for automated lattice surgery compilation and diagrammatic optimization in color code architectures.
\end{abstract}

\maketitle

\section{Introduction}
\begin{figure*}[]
    {\centering
    \includegraphics[width=\linewidth]
    {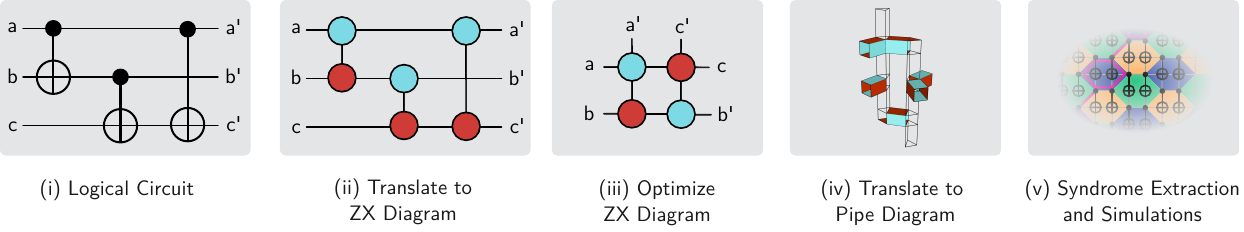}
    \par}
    \caption{Overview of steps in compilation using pipe diagrams. Starting with a logical circuit (i) it is translated into a ZX diagram (ii). A ZX diagram can be optimized in different ways and made more compact with diagrammatic reasoning (iii). Afterwards, this ZX diagram is embedded in 3D spacetime as a color code pipe diagram (iv). The distance-independent pipe diagram can be translated into specific stabilizers for given $d$ alongside its explicit correlation surfaces from which one can receive a syndrome extraction circuit and perform simulations (v). Figure (v) displays an example cutout and timestep within the syndrome extraction.}
    \label{fig:overview}
\end{figure*}

Large-scale quantum computation based on quantum error correction requires automated methods across multiple levels of abstraction. These processes are commonly referred to as \textit{compilation}.
Within the framework of topological quantum error-correcting codes and under the assumption of two-dimensional architectures, \textit{lattice surgery}~\cite{Bombin_2009_deformation, horsman_surface_2012,landahl_quantum_2014} has emerged as a standard method for implementing logical gates.

Given a logical quantum circuit in the Clifford+T gate set, compilation must be addressed on at least two distinct levels of abstraction. On the \textit{microscopic level}, one has to specify how logical qubits are encoded into patches of a given topological code and distance $d$, how lattice surgery operations are implemented, and how syndrome extraction is performed efficiently. Compilation on this microscopic level aims to construct syndrome extraction circuits automatically. On the \textit{macroscopic level}, compilation is typically formulated as a mapping and routing problem, where lattice surgery operations are scheduled to maximize parallelism of gate execution~\cite{molavi_dependency-aware_2024, beverland_surface_2022, watkins_high_2024, herzog_lattice_2025, herzog2025exploitingmovablelogicalqubits}.

Traditionally, macroscopic compilation is treated as a process in which the protocol of executing a single gate with lattice surgery is fixed in advance. Then, compilation aims to schedule these fixed building blocks as efficiently as possible, i.e., leading to a low-depth schedule. However, a more fine-grained perspective allows for greater flexibility in executing logical gates with lattice surgery, where the implementations of specific gates are not fixed beforehand but can instead be arranged and optimized in spacetime using \textit{pipe diagrams}~\cite{fowler2025surfacecode} as an intermediate representation.
Pipe diagrams provide a powerful abstraction for representing and optimizing lattice surgery protocols and have been extensively developed in the context of the surface code, supported by tools such as \texttt{tqec}~\cite{tqec}.

Despite the prominence of the surface code~\cite{fowler_surface_2012}, it is not the only viable topological code that supports universal fault-tolerant computation on planar quantum computers through lattice surgery. In particular, the \textit{color code}~\cite{bombin_topological_2006} offers several advantageous features, including transversal implementation of all single-qubit Clifford gates and lower qubit overhead in comparison to the surface code. The decoding problem is much more demanding for the color code, although rapid progress has been made on this problem~\cite{koutsioumpas_colour_2025, Kubica2023efficientcolorcode, Chamberland_2020_decoding, gidney2023new, lee_color_2025}.

These considerations motivate the extension of pipe diagram-based compilation techniques beyond the surface code to other code families, in particular to color codes.
In this work, we propose a foundation for color code-based pipe diagrams that enable the development of automated compilation methods on the microscopic and macroscopic levels. Specifically, we establish a distance-independent pipe diagram representation for the triangular color code on the 6.6.6 lattice, alongside with explicit constructions for stabilizers, correlation surfaces and syndrome extraction given arbitrary distance $d$. We present examples of macroscopic and microscopic compilation that highlight the potential of this approach. %

\autoref{fig:overview} displays the steps of compilation in this framework following an example of a toy logical circuit. Starting with a logical circuit (i), it is translated to a ZX diagram (ii), which is in turn optimized (iii). Then, the optimized ZX diagram (which may no longer resemble a quantum circuit) is embedded as a corresponding pipe diagram (iv), here for the color code, in spacetime with minimal overhead. This constitutes macroscopic compilation. The pipe diagram is then compiled down to a physical-qubit-level circuit along with a construction of correlation surfaces corresponding to logical observables for given distance $d$, which constitutes microscopic compilation. These circuits may then be simulated.%

The remainder of this paper is structured as follows. \autoref{sec-background} provides the necessary background on topological codes, previously established pipe diagrams for the surface code, the ZX calculus, and related work. We proceed with the definition of lattice surgery between color code patches and correlation surfaces and how these microscopic definitions can be represented as distance-independent pipe diagrams that can, in turn, be represented as ZX diagrams in \autoref{sec-pipe-diagrams-cc}. This is followed by a discussion of examples for macroscopic compilation in \autoref{sec-macroscopic-compilation} that highlight the benefits of pipe diagrams in general and for the color code specifically. Afterwards, we discuss how to perform syndrome extraction in this setting and showcase proof-of-principle simulation results in \autoref{sec-microscopic-compilation}. We conclude in \autoref{sec-conclusion}.

\section{Background}\label{sec-background}
\subsection{Related Work}\label{sec-related-work}

A large body of work on lattice surgery compilation follows a paradigm of fixed logical gate realizations, in which lattice surgery primitives for logical operations -- such as CNOT gates or $T$-state injections -- are defined beforehand. Compilation then amounts to scheduling these predefined building blocks in order to obtain a low-depth implementation of a given logical quantum circuit. In this context, the surface code has emerged as the predominant platform for studying lattice surgery compilation~\cite{molavi_dependency-aware_2024, beverland_surface_2022, silva_multi-qubit_2024, herr_lattice_2017, zhu_ecmas_2023, watkins_high_2024}. Many of these approaches can be understood as solving mapping and routing problems on planar graphs, where the goal is to realize the required logical circuit in low depth.

An alternative perspective is provided by \textit{pipe diagram} approaches~\cite{tqec, zhou2026topolslatticesurgerycompilation, topologiq, tan_sat_2024}, which allow for a more flexible and fine-grained description of lattice surgery computations. Rather than relying on a fixed set of predefined lattice surgery constructions for individual gates, pipe diagrams describe computation through the consistent flow of Pauli operators across a spacetime structure. This allows to reduce spacetime volume by focusing on the topological structure of the computation instead of plugging together predefined gate instructions. 
This viewpoint enables greater freedom in arranging operations and can lead to improved spacetime volume. This is also due to the fact that pipe diagrams can be translated to ZX diagrams and in ZX diagrams a notion of spatial and temporal direction does not exist -- thus, at the level of ZX diagrams, there is a lot of flexibility in deforming the structure and hence, also a great deal of flexibility for the resulting pipe diagram.
Pipe diagram methods have been studied extensively in the context of the surface code, with tools such as \texttt{tqec}~\cite{tqec} providing concrete realizations of this paradigm. 

In contrast, comparatively few works address lattice surgery compilation for the color code. Existing approaches in this setting predominantly follow the paradigm of fixed logical gate operations~\cite{herzog_lattice_2025, herzog2025exploitingmovablelogicalqubits}.
Additional ideas related to lattice surgery and compilation for color codes appear in~\cite{thomsen_low-overhead_2024, litinski_combining_2017}. However, they do not constitute automated compilation frameworks.
Consequently, there exists a gap in the literature in extending the pipe diagram representation from the surface code to color code architectures.

\subsection{Surface Code and Color Code}\label{sec-background-sc-cc}

\begin{figure}[t]
    \centering

    \subfloat[\label{fig:subfig_a_surfacecoded7}]{
        \includegraphics[width=0.47\linewidth]{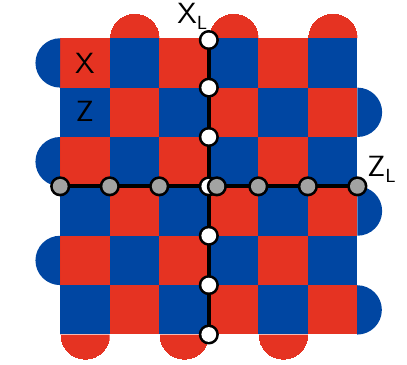}
    }
    \hfill
    \subfloat[\label{fig:subfig_b_colorcoded7}]{
        \includegraphics[width=0.47\linewidth]{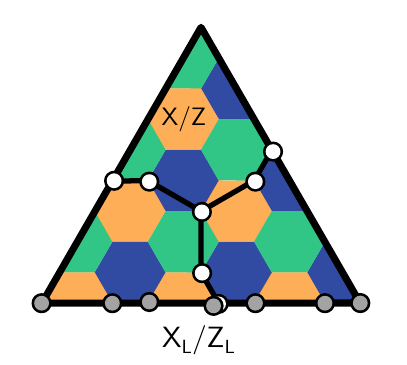}
    }

    \caption{
    (a) $d=7$ rotated surface code, where $X$ stabilizers are red faces and $Z$ stabilizers are blue faces. A representative of the $X_L$ and $Z_L$ are displayed with white and gray nodes respectively. Note that the logical operators are perpendicular to each other.
    (b) $d=7$ color code on a hexagonal tiling with data qubits on the vertices. Each face defines both an $X$ and $Z$ stabilizer on the incident qubits. Both $X_L$ and $Z_L$ logical operators are supported on the boundaries of the triangular region, with an example displayed with gray nodes. An instance of an equivalent ``star-shaped'' $X_L$ or $Z_L$ operator is displayed by white nodes.
    }
    \label{fig:combined}
\end{figure}
Topological quantum error-correcting codes~\cite{bombin_topological_2006, bombin_topological_2010, bombin_introduction_2013} are defined by placing qubits on a manifold --  in what follows, we consider the two-dimensional setting. Stabilizers~\cite{gottesman1997stabilizercodesquantumerror} of the code are associated with faces and data qubits with vertices on a 2D tiling. Most importantly, stabilizers are local objects on that tiling, while logical information is encoded in a non-local manner. This locality makes such codes particularly suitable for quantum hardware with nearest-neighbor planar connectivity, such as superconducting quantum computers. 

The \textit{surface code}~\cite{Kitaev_2003, Dennis_2002, fowler_surface_2012} is the most extensively studied example of a topological code. It is defined on a two-dimensional lattice with local X-type and Z-type stabilizers that are usually placed in a checkerboard pattern. Logical operators on a logical patch are string-like operators that connect opposite boundaries of the patch. An example of the rotated surface code for $d=7$ is displayed in \autoref{fig:combined}{a}.
In the surface code, logical $X_L$ and $Z_L$ operators are supported on strings that run along \textit{different, perpendicular directions}. This geometric asymmetry directly constrains how logical qubits can be connected and how multi-qubit operations are implemented.

The \textit{color code}~\cite{bombin_topological_2006, kubica_abc_2018} is another prominent topological code, defined on a trivalent, three-colorable lattice in which each face supports both X-type and Z-type stabilizers. An example of the triangular color code on the 6.6.6 lattice for $d=7$ is shown in \autoref{fig:combined}{b}.
In particular for color codes, logical $X_L$ and $Z_L$ operators can be supported on the same data qubits. Due to the structure of the lattice, logical Pauli operators are available in all directions, rather than being restricted to perpendicular orientations. Logical operators can also be represented with support on a `star-shaped' pattern as shown in white in \autoref{fig:combined}{b}, which respects the rotational symmetry of the code patch \footnote{Note, however, in general, a `star-shaped' logical operator does not necessarily require a strict symmetry of $2\pi/3$ rotations. The constructions of this work do not explicitly rely on a strict rotational symmetry.}. In general, one can consider a logical operator as `star-shaped' if it contains Paulis on all three sides of the code patch that are not placed on the corners of the patch. These Pauli operators constitute the endpoints of three strings that join in a point in the bulk. A special case is $d=3$ where the `star-shaped' operator consists of Pauli operators at the middle of each boundary only.

This difference in logical operator structure has important implications for logical compilation using these codes. While the surface code imposes more geometric constraints on logical connectivity and thus routing, the color code enables more flexible connections between logical qubits. Nevertheless, the spatial connectivity of logical color code patches of this kind is described by a 3-valent graph -- as each patch has three sides -- while the surface code's spatial connectivity is described by a 4-valent graph -- as each patch has four sides. Furthermore, the color code also offers reduced data qubit overhead compared to the surface code.
Most importantly, the color code admits transversal implementations of single-qubit Clifford gates, such as the $H$ and $S$ gates. In contrast, the surface code requires code deformation~\cite{fowler2019lowoverheadquantumcomputation} for the $H$ gate and injection for the $S$ gate~\cite{fowler_surface_2012, Gidney2024inplaceaccessto}. Thus, gates that must be performed physically in the surface code, can be performed in software in the color code~\cite{beverland_cost_2021} or via transversal physical implementation, which can be done much more quickly than deformation and injection, as no repeated rounds of syndrome extraction are required. Considering the application of general Clifford gates, the color code thus has potential for reduced spacetime overhead in comparison to the surface code.%

\subsection{Pipe Diagrams for the Surface Code}\label{sec-background-sc-pipes}
\begin{figure}[t]
    \centering
    \includegraphics[width=\linewidth]{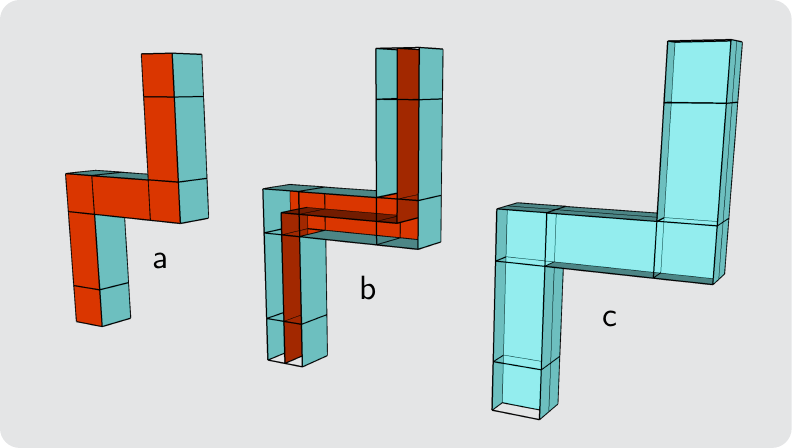}
    \caption{Surface code pipe diagram representing the movement of a logical qubit (a). X-type walls and correlation surfaces are shown in red (dark), and Z-type are shown in blue (light). For better visibility, the front walls are removed from the figure to display the X-type correlation surface (b) and the Z-type correlation surface (c).}
    \label{fig:background-tele-sc}
\end{figure}
Pipe diagrams are a well-established, distance-independent representation that is used throughout the lattice surgery-based literature for surface codes, with their modern form first displayed in \cite{Gidney_2019_MSD}.
This section only briefly summarizes the idea of pipe diagrams, while the reader should consult, e.g., Ref.~\cite{fowler2025surfacecode} for more details.
A memory experiment of a surface code can be depicted as a cube, where the vertical height corresponds to $d$ rounds of stabilizer measurements in time. The horizontal walls represent initialization and measurement bases (blue for $Z$ and red for $X$) while vertical walls represent the 1D boundaries of the surface code patch in time. 
One can consider \autoref{fig:combined}{a} as the top-down view on such a cube, where the boundaries in 2D constitute vertical walls in 3D which are colored by the color of the adjacent plaquettes, i.e., if a $X_L$ can end at a boundary, the corresponding 3D wall is red and vice versa.

An example of a simple pipe diagram in the surface code is displayed in \autoref{fig:background-tele-sc}. All three objects display the same operation with correlation surfaces explicitly shown in (b) and (c). On the bottom left cube of the 3D object, the computation starts with $d$ rounds of syndrome extraction. It is connected by a vertical pipe to another cube above it. The pipes in the pipe diagram indicate that neighboring cubes are connected together rather than both terminating at opposite faces. Pipes do not occupy a physical volume and are only drawn as if they do for better visibility. As a next step, the current patch is merged with a newly initialized patch on the right, and subsequently the `old' patch is measured out, which is indicated by the top horizontal walls. This constitutes a movement to a new position. As no logical operation is performed, this can also be considered a deformed memory experiment.
Throughout this process, the walls and their colors represent the behaviour of logical operators.

Most importantly, lattice surgery induces fundamentally random outcomes (in the absence of errors) that must be tracked and taken into account for the final outcomes. This tracking is covered by \textit{correlation surfaces} which represent the flow of Pauli operators through the structure, which is shown in \autoref{fig:background-tele-sc}{b,c}. Along those correlation surfaces, one has to track the parities of measurement outcomes, $\lambda_X$ (along the red surface) and $\lambda_Z$ (along the blue surface), and adjust the final logical operators accordingly, i.e., $X'_L = (-1)^{\lambda_X}X_L$ and $Z'_L = (-1)^{\lambda_Z}Z_L$.

Note that tracking of correlation surfaces is, in principle, a code-agnostic concept that emerges due to lattice surgery. Thus, only the specific shape of the correlation surfaces emerges from the structure of logical operators in the given topological code and will be different for color codes.

\subsection{ZX Calculus}
\begin{figure}[t]
    \centering
    \includegraphics[width=\linewidth]{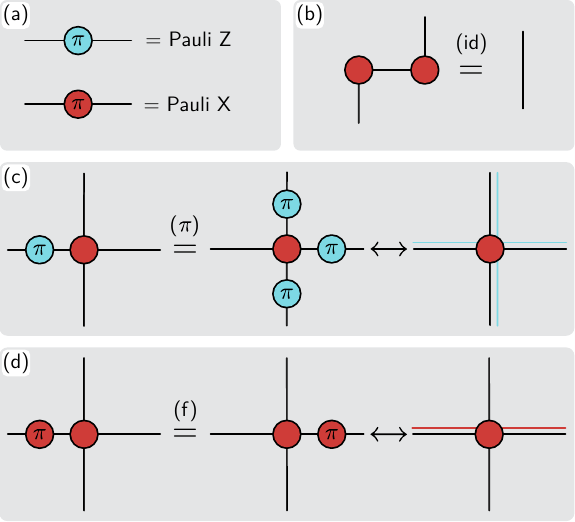}
    \caption{ZX rules for pipe diagrams. (a) Representation of Pauli $X$ and $Z$ matrices within the ZX calculus. Blue (light) nodes are Z spiders, and red (dark) nodes are X spiders. (b) ZX diagram of \autoref{fig:background-tele-sc} with open ports. (c,d) Propagation of Pauli operators through a red rank-4 junction. ZX Rules follow the labels as established in~\cite{wetering_zx-calculus_2020}.}
    \label{fig:background-zx}
\end{figure}
The ZX calculus~\cite{wetering_zx-calculus_2020, KissingerWetering2024Book} can be viewed as a useful language for lattice surgery~\cite{de_beaudrap_zx_2020}, i.e., each pipe diagram can be translated into a ZX diagram that can be altered according to the rules of the ZX calculus -- such alterations can yield a reduction in spacetime overhead for the final pipe diagram.

Furthermore, ZX diagrams are an alternative representation that guides one to observe the Pauli flow~\cite{PRXQuantum_xyz_pauliflow} through the structure, as Pauli matrices can be represented within that framework as shown in \autoref{fig:background-zx}{a}. Considering the pipe diagram from \autoref{fig:background-tele-sc} with open ports (i.e., no specified initialization and measurement procedure at the temporal beginning and end of the diagram), it can be represented as a ZX diagram as displayed in \autoref{fig:background-zx}{b}. Pushing a Pauli $X$ and $Z$ operator through this structure via $\pi$-copy ($\pi$) and spider fusion ($f$) rules yields the same correlation surfaces as displayed in \autoref{fig:background-tele-sc}{b,c}. Moreover, utilizing the identity removal ($id$) rule shows that this pipe diagram was nothing more than an identity gate -- i.e., despite of moving the patch in space, no logical operation was applied.

While this example displays simple correlation surfaces, the behaviour becomes more intricate for more complex structures. Considering a single red phaseless ZX node with four legs as in \autoref{fig:background-zx}{c,d}, one can push through the Pauli $Z$ and $X$ operators accordingly. For a Pauli of the opposite color (\autoref{fig:background-zx}{c}), the Pauli ``spreads'', while for a Pauli of the same color (\autoref{fig:background-zx}{d}) the Pauli can ``decide'' which way it goes. This behaviour is indicated in a more abstracted representation on the right, where blue and red lines indicate the behaviour of the Pauli operators. Both figures work exactly the same way with node colors interchanged.

Note that the ZX diagram is a code-independent view, and thus the same rules apply for the surface code and color code. However, the codes differ in how many legs can be directly embedded for a single node due to different geometrical constraints. Note that a general ZX-diagram may not be directly embedded in spacetime, e.g., if the degree of a node is too high. In such cases, it may be necessary to add nodes and edges to enable the compilation to a pipe diagram.

\section{Pipe Diagrams for the Color Code}\label{sec-pipe-diagrams-cc}

\subsection{Standalone Prism}
\begin{figure}[t]
    \centering
    \includegraphics[width=0.5\linewidth]{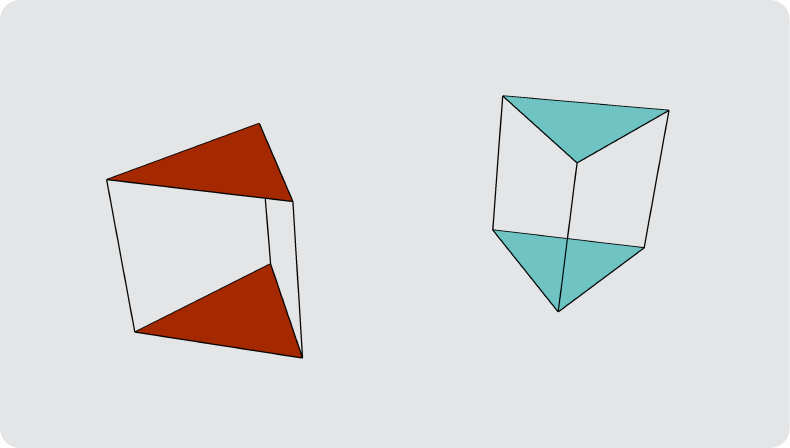}
    \caption{Standalone prisms that represent separate memory experiments of a single patch with $d$ rounds of syndrome extraction. The color of the horizontal walls determines the initialization and measurement bases. Red (dark) indicates initialization of data qubits in $|+\rangle$ and measurement in the $X$ basis. Blue (light) indicates initialization in $|0\rangle$ and measurement in the $Z$ basis. As $X_L$ and $Z_L$ representatives can both terminate on vertical walls, vertical walls do not receive a color assignment.}
    \label{fig:standalone-prism}
\end{figure}
A memory experiment of a single color code patch (\autoref{fig:combined}{b}) can be represented in a distance-independent way by a prism with an equilateral triangular base (\autoref{fig:standalone-prism}). Similarly to the surface code pipe diagrams, the extension into the vertical direction represents $d$ rounds of syndrome extraction, given a distance $d$ color code patch. 

The colors of the horizontal faces indicate the preparation and measurement bases, respectively. If time flows from bottom to top, the lower horizontal face indicates the preparation basis, i.e. red (dark) for $|+\rangle$ preparation of data qubits while blue (light) stands for $|0\rangle$ preparation of data qubits. In turn, the top horizontal face displays the measurement basis -- red for an $X$-basis measurement of the data qubits and blue for $Z$.

In contrast to the surface code pipe diagrams, vertical walls do not have colors in the color code case. This is due to the fundamentally different structures of logical operators between surface and color codes. While the surface code has perpendicular $X_L$ and $Z_L$ operators, which terminate at defined boundaries of the patch and necessitate colorings of vertical walls; the color code offers a more flexible logical operator structure: Both $X_L$ and $Z_L$ operators can be placed along any of the three boundaries (\autoref{sec-background-sc-cc}). Thus, they are less restricted, which makes a coloring of vertical walls superfluous for a single patch in spacetime.

\subsection{Vertical Pipe}
\begin{figure}[t]
    \centering
    \includegraphics[width=\linewidth]{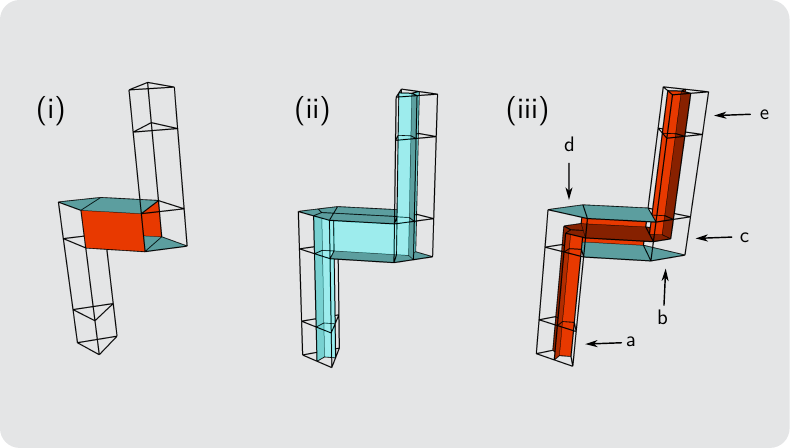}
    \caption{Pipe diagram representing a logical movement in the color code (i). For better visibility, the front wall is removed from the figure to display the Z-type correlation surface (ii) and the X-type correlation surface (iii).}
    \label{fig:teleportation}
\end{figure}
To represent logical computations with pipe diagrams, prisms must be connected by pipes as can be seen in the example in \autoref{fig:teleportation}. To connect prisms along the time axis, vertical pipes are added. For the same reasoning as above, the vertical walls do not receive a color. However, an adaption to the representation of vertical pipes is applied as soon as logical single-qubit Clifford gates are applied (\autoref{subsec-full-clifford}).

Since logical operators in the color code can take so many different forms, there is also freedom to choose the representation of the logical operator through time, i.e., the vertical correlation surface in a vertical pipe. The present work adopts ``star-shaped'' (\autoref{sec-background-sc-cc}) logical operators as the default choice on both the microscopic level and for the distance-agnostic representation.

\subsection{Horizontal Pipe}~\label{subsec-hor-pipe}
Choosing the ``star-shaped'' operator representatives as the default representation is particularly important as soon as horizontal pipes come into play. Choosing ``star-shaped'' operators is advisable as they allow for straightforward construction of correlation surfaces compared to other logical operator choices, which would require vertical correlation surfaces (in horizontal pipes) being constructed on a case-by-case basis. For instance, choosing one of the logical operators along a boundary of the color code patch would lead to string operators through the structure with peculiar shape that have to be constructed on a case-by-case basis due to the involved domain walls in space. Furthermore, ``star-shaped'' logical operators allow for a representation in the distance-agnostic pipe diagram, which is rotation symmetric such that drawing the distance-independent correlation surfaces becomes particularly elegant. This is detailed in the following by explaining the choice of lattice surgery scheme.

\begin{figure}[t]
    \centering
    \includegraphics[width=0.8\linewidth]{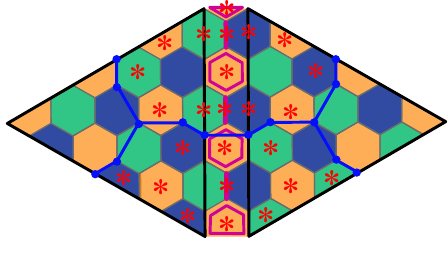}
    \caption{Merge of two $d=7$ color code patches. Plaquettes with pink outlines represent single-type $X$ stabilizers. Pink lines represent weight-2 $X$-type stabilizers. Any other plaquette is both an $X$ and $Z$ type stabilizer generator. How the $Z_L$ operator can be extended is displayed by the dark blue structure. The $X$ stabilizer product representing $X_LX_L$ of the previously split patches is displayed by red asterisks.}
    \label{fig:merge2patch}
\end{figure}
Considering lattice surgery between logical patches via semi-transparent domain walls (STDWs)~\cite{kesselring_anyon_2024}, horizontal pipes in a pipe diagram convey information about the initialization and measurement of auxiliary data qubits between the patches as well as information about the logical operators -- merging multiple logical color code patches breaks the otherwise existing identical support of $X_L$ and $Z_L$ operators. This section explains how lattice surgery is done in this work on the level of stabilizers and how the pipe diagram picture emerges from it.

Consider \autoref{fig:merge2patch} in which two $d=7$ color code patches are merged into one joint code. The plaquettes defined by the 6.6.6 lattice determine the $X$ and $Z$ stabilizers. However, those plaquettes at the interface of both patches with pink outlines define single-type stabilizers -- in this example, $X$ stabilizers. Note that weight-2 $X$-type stabilizers are indicated by pink lines. These single-type stabilizers define an STDW~\cite{kesselring_anyon_2024} that connects the patches in a proper manner. The domain walls dictate how string operators can pass through it -- in this case, orange $Z$-type anyons can pass through the domain wall, while blue and green such anyons cannot pass through it. Orange $X$-type anyons condense in the domain wall and it is transparent for green and blue such anyons.

Note that lattice surgery for color codes can be done in a variety of ways~\cite{kesselring_anyon_2024, landahl_quantum_2014, thomsen_low-overhead_2024}. We choose this specific technique because it does not require stabilizers with weight $>6$ and it allows placing an arbitrary number of triangles on the 6.6.6 lattice without the need to deform the substrate~\cite{herzog_lattice_2025}. Furthermore, it allows to do $Z_LZ_L$ and $X_LX_L$ measurements separately based on the specific choice of single-type stabilizers in the STDW. One could imagine working with fully transparent domain walls as well in order to measure $X_LX_L$ and $Z_LZ_L$ simultaneously. However, such a scheme would not give us the control over the choice of logical operators that is needed to realise logical computation based on the ZX picture (\autoref{subsec-zx-junctions}).

Due to the aforementioned interplay between anyons and the domain wall, it is desirable to choose ``star-shaped'' $Z_L$ representatives to extend over multiple patches (indicated in dark blue in \autoref{fig:merge2patch}) as these guarantee the correct anyon type available at the domain wall. Note that in this setting, a $Z_L$ operator is forced to extend throughout the whole merged structure as its anyonic excitations cannot condense at the domain wall due to its single-type $X$ stabilizers. For consistency, also $X_L$ and $X_L'$ operators at the respective patches are chosen as ``star-shaped''. However, their anyonic excitations can condense at the domain wall such that they are not forced to extend through the structure, as is the case for the $Z_L$. They propagate with a horizontal correlation surface, equivalent to an $X_LX_L$ measurement. The $X_LX_L$ measurement outcome of previously separate patches (or the parity $\lambda_X$ that changes $X_L$ to $X_L'$) can be determined by the product of merged stabilizers belonging to the set $CS_X$ indicated by red asterisks in \autoref{fig:merge2patch}. The explicit construction of those horizontal correlation surfaces is detailed in~\autoref{subsec-explicit-cs}. This specific stabilizer product has trivial action in the bulk and is nontrivial at the data qubits of the $X_L^{(')}$ representatives, as $X^{n}=\mathds{1}$ for even $n$ and $X^{n}=X$ for odd $n$. This means that each data qubit that is not part of the logical operator representative is ``touched'' an even number of times by stabilizers in $CS_X$. For a standard lattice surgery step that merely measures $X_LX_L$, one begins with two separate patches and initializes the data qubits in between the patches in $\ket{0}$. Then, the merged stabilizers are measured and subsequently split again by measuring split stabilizers and the data qubits at the interface in the $Z$ basis. An analogous construction applies if bases are interchanged with $Z$-type single-type stabilizers at the interface.

From this, the abstraction to horizontal pipes emerges. To illustrate this, consider \autoref{fig:teleportation}, which represents the logical movement of a color code patch from one position to another position. Starting with stabilizer measurements of the initial patch (a), one proceeds with the initialization of data qubits at the interface and in the second patch in the state $\ket{0}$ (b). This is followed by $d$ rounds of syndrome extraction for the merged code (c) as displayed previously in \autoref{fig:merge2patch}. Then, the initial patch and the interface qubits are measured in the $Z$ basis (d), and the movement is completed and followed in this case by another $d$ rounds of syndrome extraction of the new patch (e). 

Thus, this is naturally depicted by certain colorings of the pipe diagram's faces. The horizontal pipe has blue horizontal walls that indicate initialization and measurement in $Z$ basis. The same holds for the neighboring prisms. The vertical walls of the horizontal pipe are colored in red as a $Z_L$ operator could not terminate at this pair of walls due to the single-type $X$ stabilizers at the interface. From this, the structure of the correlation surfaces in the pipe diagram also emerges: while the $X$ and $Z$ correlation surfaces remain the same for single-patch memory, they have different structures as soon as horizontal pipes are added. The red correlation surface has to terminate at the red walls such that it horizontally connects the first and second patch, which is represented by the stabilizer product indicated by red asterisks in \autoref{fig:merge2patch}. We denote the subset of these stabilizers as $CS_X$. In turn, the $Z$ correlation surface remains fully vertical, as the $Z_L$ string operator representative has to be extended between both patches as shown in \autoref{fig:merge2patch}. Z-type anyons cannot condense at the interface; only X-type anyons can do this, thus the extension of the $Z_L$ is inevitable.

Note that the whole discussion above works analogously with $X$ and $Z$ stabilizers interchanged. Furthermore, note that this fundamental qualitative difference between horizontal and vertical pipes is a key difference between color code pipe diagrams and surface code pipe diagrams. The latter can be rotated, such that the roles of spatial and temporal pipes are interchanged -- this cannot be done in the color code case anymore due to the different geometry and different roles of vertical and horizontal pipes.

\subsection{Explicit Correlation Surfaces for Given $d$}\label{subsec-explicit-cs}
\begin{figure}[t]
    \centering
    \includegraphics[width=0.9\linewidth]{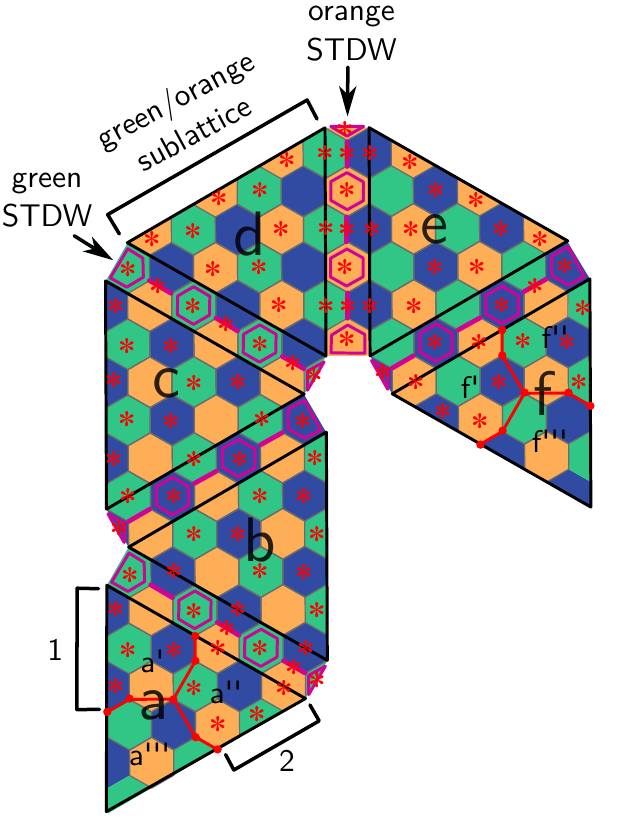}
    \caption{Horizontal $X$ correlation surface for a sequence of $d=7$ triangular color code patches. Each plaquette is an $X$ and $Z$-type stabilizer except for the plaquettes and weight-2 stabilizers marked in pink, which are only $X$-type stabilizers. Boundary patches are defined by patch $a$ and $f$, while bulk patches are defined by $b,c,d,e$.}
    \label{fig:horcs-construction}
\end{figure}

\begin{figure}[t]
    \centering
    \rotatebox{90}{\includegraphics[width=0.37\linewidth]{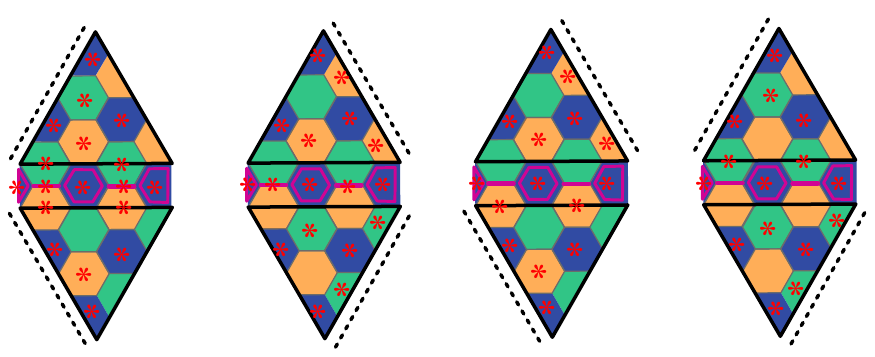}}
    \caption{Given different choices of sublattices for the stabilizer subset $CS_X$, different choices for adding stabilizers from the STDW to $CS_X$ must be made. A dotted line indicates that a triangle side is ``free'', i.e., the product of $CS_X$ must be trivial on those sides. The open side without a dotted line is assumed to be connected to another, not depicted, bulk patch. If adjacent bulk patches use different sublattices, weight-2 stabilizers have to be taken into account (left two figures). If the sublattice is chosen to be the same in the adjacent patches, weight-2 stabilizers are not needed (right two figures).}
    \label{fig:horcs-construction_2}
\end{figure}
While in the surface code case horizontal correlation surfaces, i.e. subsets of stabilizers $CS_{X/Z}$, can be chosen to be just one of the two colors in the checkerboard pattern, as X and Z stabilizers are assigned to one of these colors each -- the construction for color codes with STDWs is more intricate due to the different structure of the underlying 6.6.6 lattice, which is 3-colorable. Furthermore, the color code is self-dual such that a simple assignment as in the surface code case is not possible.

In the current setup, triangles of fixed distance $d$ tesselate the 6.6.6 lattice with connections by STDWs. Single-type stabilizers at the intersection are chosen to be either $X$- or $Z$-type depending on the type of lattice surgery that is aimed to be performed.
\textit{Vertical correlation surfaces} of connected ``star-shaped'' operators can be generated by fixing a seed ``star-shaped'' operator at some randomly chosen patch. Since the tessellation of triangular patches is governed by the triangle group -- generated by reflections along the sides of the triangles -- the seed operator can be reflected across patch boundaries, guaranteeing consistent connectivity throughout the structure. %
Reflections guarantee that the ``star-shaped'' operators are connected in such a way that anyonic properties are respected when a string operator propagates through an STDW. Note that the anyonic properties (described in \autoref{subsec-hor-pipe}) also force the operators of the vertical correlation surface to be extended throughout the whole structure as they cannot terminate at the STDW.

It is worth emphasizing that if a logical operator along a boundary of the patch was chosen as seed operator instead, such a simple reflection-based construction would not be possible, as their anyons would not automatically match the required behaviour dictated by the STDW -- a case-by-case construction would be necessary.

\textit{Horizontal correlation surfaces} require a slightly more complex construction as displayed in \autoref{fig:horcs-construction} for a set of consecutive color code patches. The generalization to connected patches that are not placed strictly consecutively is straightforward and is explained in the next paragraph. Note that one can assign each STDW a color of the 6.6.6 lattice, which is the color of the single-type stabilizers at the interface between patches. One can construct the stabilizer product constituting the $X$-type horizontal correlation surface in the following way, by distinguishing the bulk and the boundary patches:
\begin{enumerate}
\item The bulk is formed by all color code patches along the path that are supposed to have trivial action. In \autoref{fig:horcs-construction}, the patches marked with $b,c,d,e$ are bulk patches.
\item In the given structure, each bulk patch has two sides that are connected to another patch via an STDW, and one of the triangle's sides is ``free''. To choose the correct subset of stabilizers for the horizontal $X$-type correlation surface $CS_X$, one has to determine the ``color'' of the adjacent STDWs. The color of the STDW is determined by the color of the single-type plaquettes on the 3-colored 6.6.6 lattice. Consider, for instance, patch $d$. It has a green and orange STDW such that we choose the green/orange sublattice and add it to $CS_X$. This choice guarantees trivial action on the ``free'' side of the patch, as its data qubits are ``touched'' twice by stabilizers in $CS_X$.
\item After the stabilizer subset is chosen from the bulk's patches, one has to determine which stabilizers from the STDW are included in the subset as well. There are multiple possibilities, in particular if the sublattice differs between adjacent bulk patches. These possibilities are shown in \autoref{fig:horcs-construction_2}.
\item After the correct stabilizers from the bulk patches and STDWs are added to the subset, one has to take care of the two boundary patches. One can say that the star operators `split' the patch into three thirds. In \autoref{fig:horcs-construction}, one can see three sections of stabilizers in the boundary patch $a$ and $f$ indicated by $a', a'', a'''$ and $f', f'', f'''$ respectively. One needs to add stabilizers into the subset which are neighbors of an STDW, here $a',a''$ and $f', f''$. Consider patch $a$, which has two free sides as it is a boundary patch. The product of stabilizers in $CS_X$ has to act trivially everywhere except on the location of the ``star-shaped'' operator $X_L$. This can be reached by adding stabilizers from $a'$ based on the colors along the side marked with a $1$. The blue/green sublattice guarantees that the product's action is trivial on the free side but is not trivial on the star operator. The same holds for $a''$ with the orange/green sublattice based on the free side $2$.
\item Then, also the subset from the STDW adjacent to each boundary patch has to be chosen accordingly, such that trivial action is guaranteed everywhere except the star operator. For this, a mixture of \autoref{fig:horcs-construction_2} is applied.
\end{enumerate}

\begin{figure}[t]
    \centering
    \includegraphics[width=0.7\linewidth]{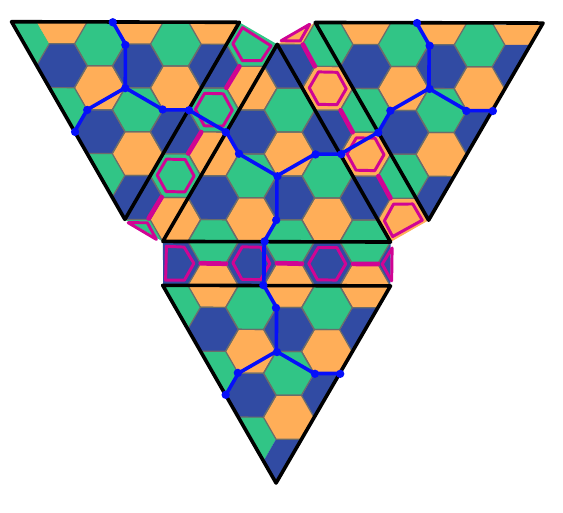}\\[0.1cm]
    {\small (a)}\\[0.2cm]
    \includegraphics[width=0.7\linewidth]{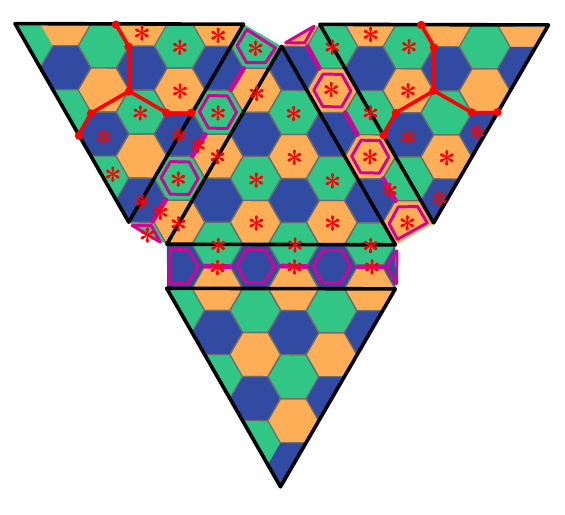}\\[0.1cm]
    {\small (b)}
    \caption{(a) ``Spreading'' of a $Z_L$ operator (=vertical correlation surface). (b) ``Choice'' of path through structure of the $X_L$ operator (=horizontal correlation surface).}
    \label{fig:rank-3-spatial}
\end{figure}

While this example had the structure of a chain of triangular color code patches, the structures can differ from such shapes as, for instance, shown in \autoref{fig:rank-3-spatial}. A structure can allow for multiple different paths of the horizontal correlation surface as shown in  \autoref{fig:rank-3-spatial}{b}. If this is the case, the discussion from above only changes slightly, as ``free'' sides have to be made trivial in the product of $CS_X$ by adding weight-2 stabilizers to the set as shown on the lower side of the middle patch.

Note that while the explicit geometric construction is instructive, the elements of $CS_X$ can also be found by solving a linear equation. Given the check matrix $H_X \in \mathbb{F}_2^{r_X\times n}$ with the number of data qubits $n$ and the number of X-type stabilizers $r_X$ as well as a vector $\mathbf{\ell}_X \in \mathbb{F}_2^n$ in which each entry is nonzero if a data qubit is part of the pair of logical operators, one can solve for the vector $\mathbf{cs}_X \in \mathbb{F}_2^{r_X}$ describing the stabilizers involved in the subset $CS_X$ following $H_X^T \, \mathbf{cs}_X = \mathbf{\ell}_X$.

\autoref{fig:rank-3-spatial}{a} displays a vertical correlation surface (blue) which is retrieved by reflections as explained above.
Note that the vertical correlation surface (a) spreads throughout the structure and the horizontal correlation surface (b) ``decides'' for a particular path. This behavior is linked to ZX diagrams that have a direct relation to such structures.

\subsection{ZX Junctions}\label{subsec-zx-junctions}
\begin{figure}[t]
    \centering
    \includegraphics[width=0.9\linewidth]{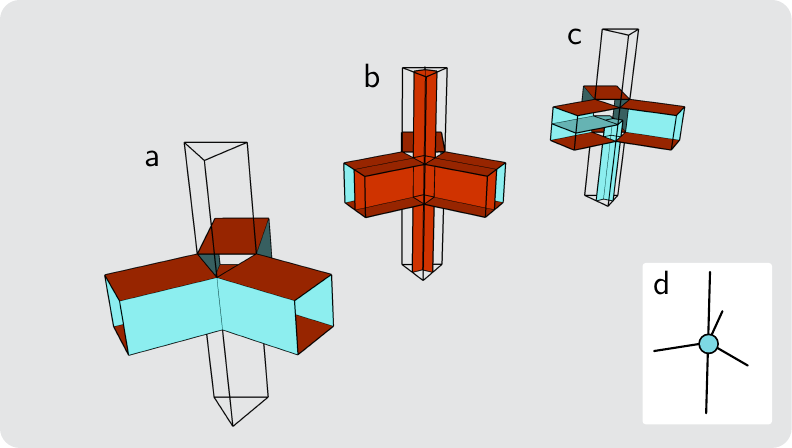}
    \caption{Pipe diagram of a rank-5 junction (a) of a blue ZX node (d). The spreading $X$ correlation surface is displayed in (b) and an example ``choice'' of the $Z$ correlation surface is displayed in (c).}
    \label{fig:rank-5-junction}
\end{figure}
The behaviour of $Z_L$ and $X_L$ in the microscopic structure in \autoref{fig:rank-3-spatial} matches what is expected from the emerging pipe diagram representation and the correspondence to ZX diagrams. Representing the middle patch with open ports both in space and time with interchanged roles of $Z$ and $X$ yields the pipe diagram shown in \autoref{fig:rank-5-junction}{a}. Exactly as in the microscopic discussion above, the vertical correlation surface ``spreads'' throughout the whole structure (\autoref{fig:rank-5-junction}{b}), while a horizontal correlation surface through spatial pipes ``decides'' on a path (\autoref{fig:rank-5-junction}{c}). This represents the logic from \autoref{fig:background-zx}, which enables a correspondence between pipe diagrams and ZX diagrams.

Thus, one can construct distance-independent correlation surfaces based on graphical rules derived from the microscopic picture that match the desired ZX structure. Whenever we encounter a prism that is connected to horizontal pipes with blue vertical walls, this is representative for a blue ZX node. In turn, a red correlation surface in the horizontal pipes has to be aligned vertically, such that it connects to the red horizontal walls (\autoref{fig:rank-5-junction}{b}). Furthermore, due to single-type $Z$ stabilizers in the STDWs, the $X$ correlation surface cannot terminate at any of the horizontal pipes and therefore ``spreads'' through the whole structure. As a vertical correlation surface cannot simply terminate at an arbitrary point, it also extends through the top and bottom vertical pipes.
In contrast, the blue correlation surface has to ``decide'' on a path with one example displayed in \autoref{fig:rank-5-junction}{c}. This is because this correlation surface is forced to be horizontal within the horizontal pipe as it has to terminate in the blue vertical walls. Alternatively to the displayed example, it is, for instance, also possible to extend a purely vertical blue correlation surface through the vertical pipes without going through the horizontal pipes. This is possible because in this setup, the $Z_L$ star operator can terminate at all the neighboring domain walls due to the involved single-type $Z$ stabilizers. This discussion holds for interchanged colors as well.

 If a prism terminates the diagram at the end of a temporal pipe, the ZX node color is uniquely defined by the opposite color of the measurement/initialization of the prism; if a prism is connected by two temporal pipes only, the ZX node's color is ambiguous. %
 Finally, if the prism connects to at least one horizontal pipe in the bulk, its color follows that of the vertical walls of the horizontal pipes.
 
\begin{figure}[t]
    \centering
    \includegraphics[width=0.9\linewidth]{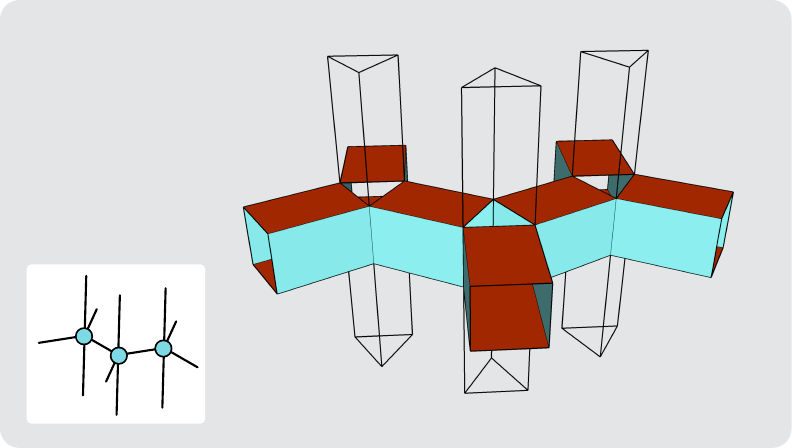}
    \includegraphics[width=0.9\linewidth]{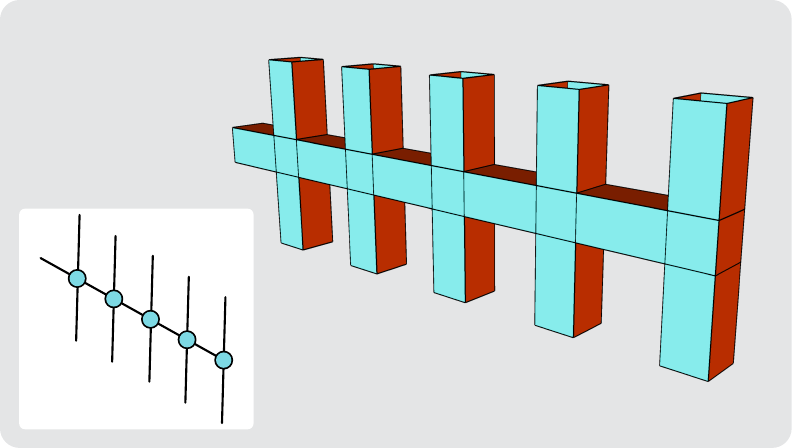}
    \caption{Multiple junctions with a total of $n=11$ open legs as pipe diagram of the color code (top) and the same ZX diagram embedded in the surface code (bottom).}
    \label{fig:multiple-junctions.}
\end{figure}

Finally, the structure of the color code allows ZX nodes with degree up to 5 to be directly embedded in spacetime. For the surface code, one can directly embed a degree-4 node. Higher-degree nodes need to be split up into multiple junctions in spacetime. To embed a degree $n$ ZX junction in 3D spacetime, there are $k_{sc} = \lceil \frac{n-2}{2} \rceil$ surface code junctions necessary. For the color code however there are $k_{cc}=\lceil \frac{n-2}{3} \rceil$ junctions necessary. Thus, \mbox{$k_{sc}\geq k_{cc}$}. An example for $n=11$ is shown for both codes in \autoref{fig:multiple-junctions.}. Overall, this means that the spatiotemporal connectivity of a color code is described by a 5-valent graph, while it is a 4-valent graph for the surface code.

The ZX representation is the key for macroscopic compilation with pipe diagrams as the ZX representation allows for optimizations and thus a low spacetime overhead embedding in spacetime can be sought.

\subsection{Full Clifford Gate Set}\label{subsec-full-clifford}
In the next section (\autoref{subsec-cnot}) we will show that such a lattice surgery architecture allows us to recreate the logical CNOT gate. In addition, the color code has the feature that any Clifford gate on a single logical qubit can be applied transversally. Together with the CNOT, this enables universal Clifford computation.

Explicitly, the single-qubit Clifford group is generated by the $H$ and $S$ gates; the logical $H$ is equivalent to applying a physical Hadamard on each data qubit, while the logical $S$ gate is implemented by a combination of physical $S$ and $S^\dagger$ gates. It is possible to find a bicoloration of the data qubits on the color code patch. %
The bicoloration partitions the data qubits into two sets; the set that includes the corner qubits of the patch receives an $S$ gate, while the other set receives an $S^\dagger$ gate~\cite{litinski_combining_2017, kubica_universal_2015}.

\begin{figure}[t]
    \centering
    \includegraphics[width=0.7\linewidth]{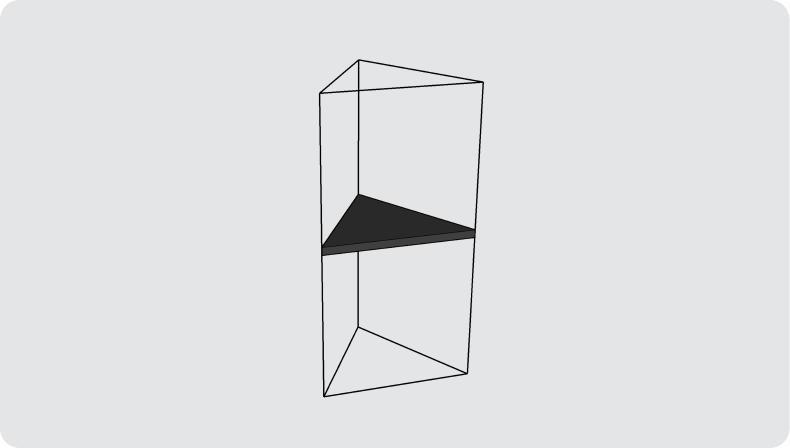}
    \caption{Transversal $S$ and $H$ gates are displayed as black plane in a pipe. The single-qubit gates are assumed to be applied in the last or first round of the prisms it connects.}
    \label{fig:h_s_gate}
\end{figure}

Under the assumption that arbitrary physical single-qubit gates can be applied on the hardware architecture, one can assume that a sequence of $S, S^\dagger$ and $H$ can be directly applied to some data qubit. Therefore, given any prism of $d$ rounds of syndrome extraction, one can add an arbitrary logical Clifford gate within the very first or last of the $d$ rounds of syndrome measurements. This comes at an insignificant spacetime cost and does not generate correlated errors on data qubits. This can be represented as a black plane in a pipe as shown in \autoref{fig:h_s_gate}, where one can consider the physical single-qubit gates to be applied either in the first or last round of syndrome extraction of the prisms this pipe connects.

Note that this is a significant benefit of the color code in comparison to the surface code, where for logical Hadamard gates code deformation~\cite{Bombin_2009_deformation, vuillot_code_2019} is required in general and $S$ gates require complex $Y_L$ state preparation or $Y$-basis measurement circuits~\cite{Gidney2024inplaceaccessto}.

\begin{figure}[t]
    \centering
    \includegraphics[width=0.9\linewidth]{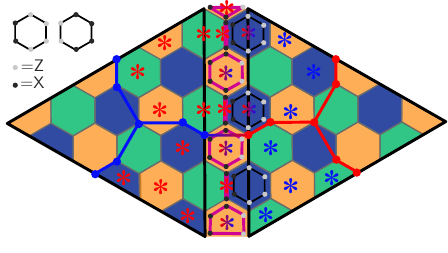}
    \caption{Execution of a spatial Hadamard gate, that swaps the roles of $X$ and $Z$-type correlation surface requires mixed-type stabilizers. The single-type stabilizers in the interface with pink outlines are marked with additional dots that indicate a $Z$ (light grey) or an $X$ (dark grey) at the data qubits. To make this a valid stabilizer code, additional mixed-type stabilizers are necessary, indicated by a black outline. On each of those locations, two stabilizers are placed: The one shown in the figure, as well as one with interchanged roles of $X$ and $Z$, as indicated at the top left corner of the figure. This stabilizer structure causes the horizontal correlation surface (indicated by red, blue, and mixed asterisks) to change color, as well as the vertical correlation surface (indicated by the string operator) to change color as well.}
    \label{fig:spatial_hadamard}
\end{figure}

Even though we will stick to temporal execution of single-qubit Clifford gates in this paper, it should be noted that it is possible to perform e.g. a spatial Hadamard nevertheless in the color code as shown in \autoref{fig:spatial_hadamard}. Such a procedure maps the $X$ operator to a $Z$ and vice versa. To achieve this, mixed-type stabilizers are required as shown in the figure: The single-type stabilizers (pink outline) are mixed $X$ and $Z$ stabilizer indicated by the dark grey (X) and light grey (Z) dots. The double-type stabilizers with black outlines are adapted to two mixed-type stabilizers as shown in the top left corner of the figure. 

In a similar manner, one can apply an $S$ gate as well by redefining the mixed-type stabilizers. An $S$ gate transforms Paulis as $SXS^\dagger = Y$ and $SZS^\dagger = Z$ which has to be reflected in the behavior of the logical operators. The single-type stabilizers with pink outlines are chosen to have $X$ Pauli operators on dark grey nodes and $Y$ Pauli operators on light grey nodes. In turn, the double-type stabilizers with black outlines contain a pure $Z$ stabilizer as well as a mixed-type stabilizer with $X$ on dark grey and $Y$ on light grey nodes. %
This transforms the horizontal correlation surface to turn from $X$ to $Y$, and the vertical $Z$ correlation surface remains as it is. %
However, it is not apparent how a reliable ZX correspondence can be established with spatial $S$ gates, which is why we stick to general single-qubit Cliffords executed in the temporal direction.


\section{Macroscopic Compilation}\label{sec-macroscopic-compilation}
Having defined the representation of pipe diagrams for the color code, one can utilize it and perform logical, macroscopic compilation. This section covers multiple examples of how low spacetime volume representations of specific logical operations can be found. We start with the CNOT gate in~\autoref{subsec-cnot} and continue with a dense representation of three CNOT gates in~\autoref{subsec-3cnots}. We finish with a specific comparison between the surface code and color code pipe diagrams in~\autoref{subsec-compare-sc-cc} as well as a summary of the benefits and deficits of the color code in~\autoref{subsec-benefits-cc}.
\subsection{CNOT Gate}\label{subsec-cnot}
\begin{figure}[t]
    \centering
    \includegraphics[width=\linewidth]{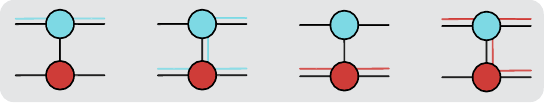}
    \caption{The CNOT gate in terms of a ZX diagram together with four different correlation surfaces.}
    \label{fig:zx_cnot_cs}
\end{figure}
\begin{figure}[t]
    \centering
    \includegraphics[width=\linewidth]{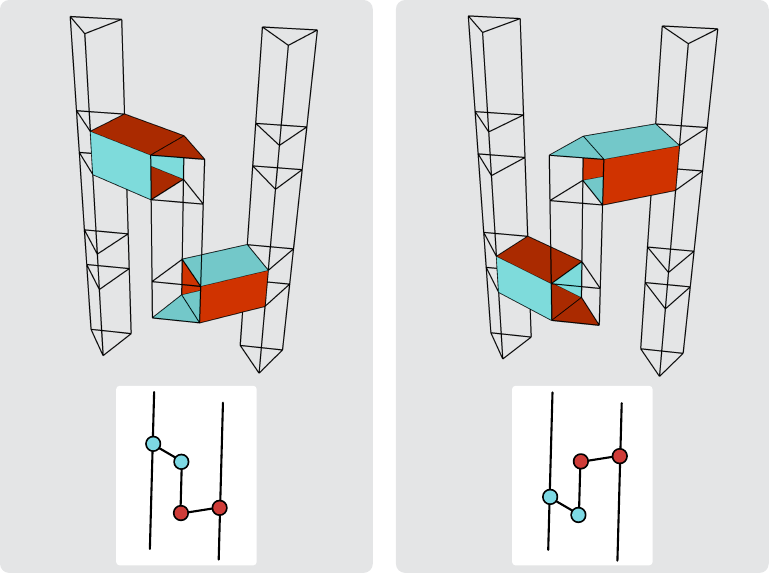}
    \caption{Two variations of the CNOT gate by embedding into pipe diagram via adding ZX nodes with the (id) rule.}
    \label{fig:pipes_cnot_nocs}
\end{figure}
To create a pipe diagram that represents a CNOT gate, one starts with its representation in the ZX calculus as shown in \autoref{fig:zx_cnot_cs}. The aim is to find a pipe diagram that has low spacetime overhead and reflects the same behaviour of the Pauli operators.

At first, we aim for a representation that has temporal open ports only, i.e., a diagram with standard time ordering. With this in mind, one has to add two more identity nodes to the ZX diagram. Two equivalent constructions for corresponding pipe diagrams are possible, as shown in \autoref{fig:pipes_cnot_nocs}. By explicitly drawing the correlation surfaces in the pipe diagram following the logic of the propagation of horizontal and vertical correlation surfaces (not shown here), one can observe exactly the same behavior as in the original ZX diagram of the CNOT gate. Furthermore, this structure directly resembles the typical measurement-based scheme for the CNOT gate~\cite{horsman_surface_2012, landahl_quantum_2014} as each horizontal pipe can be considered as a $Z_LZ_L$ or $X_LX_L$ measurement.

As drawn in \autoref{fig:pipes_cnot_nocs} (left), the structure occupies a spacetime volume given by three prisms in space ($s=3$) and two prisms in time ($t=2$), giving $STV_\text{naive} = s \cdot t = 6$. In this section, however, we adopt a less conservative count: the prisms in the lower left and upper right, which are not connected to horizontal pipes, are not strictly necessary to reproduce the logic of the computation, so we exclude them from the count. They are just identities entering the open ports. This gives an effective spacetime volume of $STV = 4$ for the CNOT gate representation.

\begin{figure}
    \centering
    \includegraphics[width=0.9\linewidth]{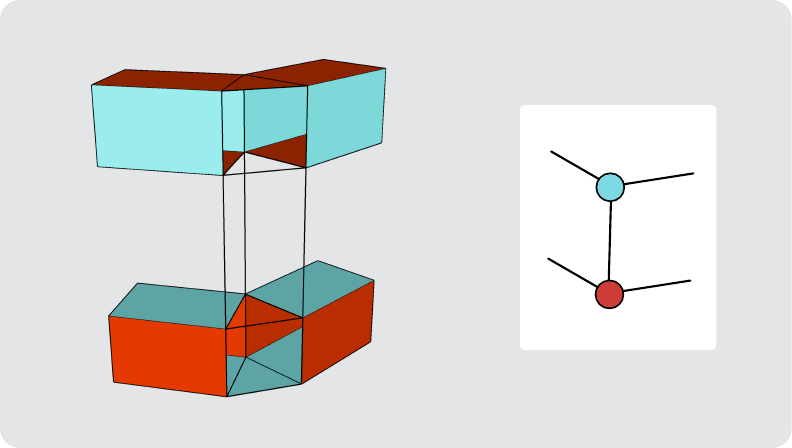}
    \caption{By allowing open spatial ports, one can create a CNOT gate pipe diagram with only two involved nodes.}
    \label{fig:cnot_spatial_ports}
\end{figure}

However, by lifting the constraint of having temporal open ports only, one can allow spatial open ports, which further reduces the overhead down to $STV=2$ as displayed in \autoref{fig:cnot_spatial_ports}. Again, correlation surfaces can be drawn in the pipe diagram, which represent the structure of the respective ZX diagram. Thus, lifting the usual notion of space and time helps to reduce the overall spacetime volume~\footnote{A few words on simulating such structures are in order here. Simulations of pipe diagrams require a deterministic observable, i.e., a correlation surface with deterministic measurement outcomes. To achieve this, one usually adds prisms/cubes at the open ports of a pipe diagram. While diagrams ending temporally are straightforward to simulate, diagrams ending spatially require a more careful construction of the correlation surface such that a deterministic measurement result can be guaranteed.}.

\subsection{Multiple CNOT Gates}\label{subsec-3cnots}
\begin{figure}[t]
    \centering
    \includegraphics[width=0.9\linewidth]{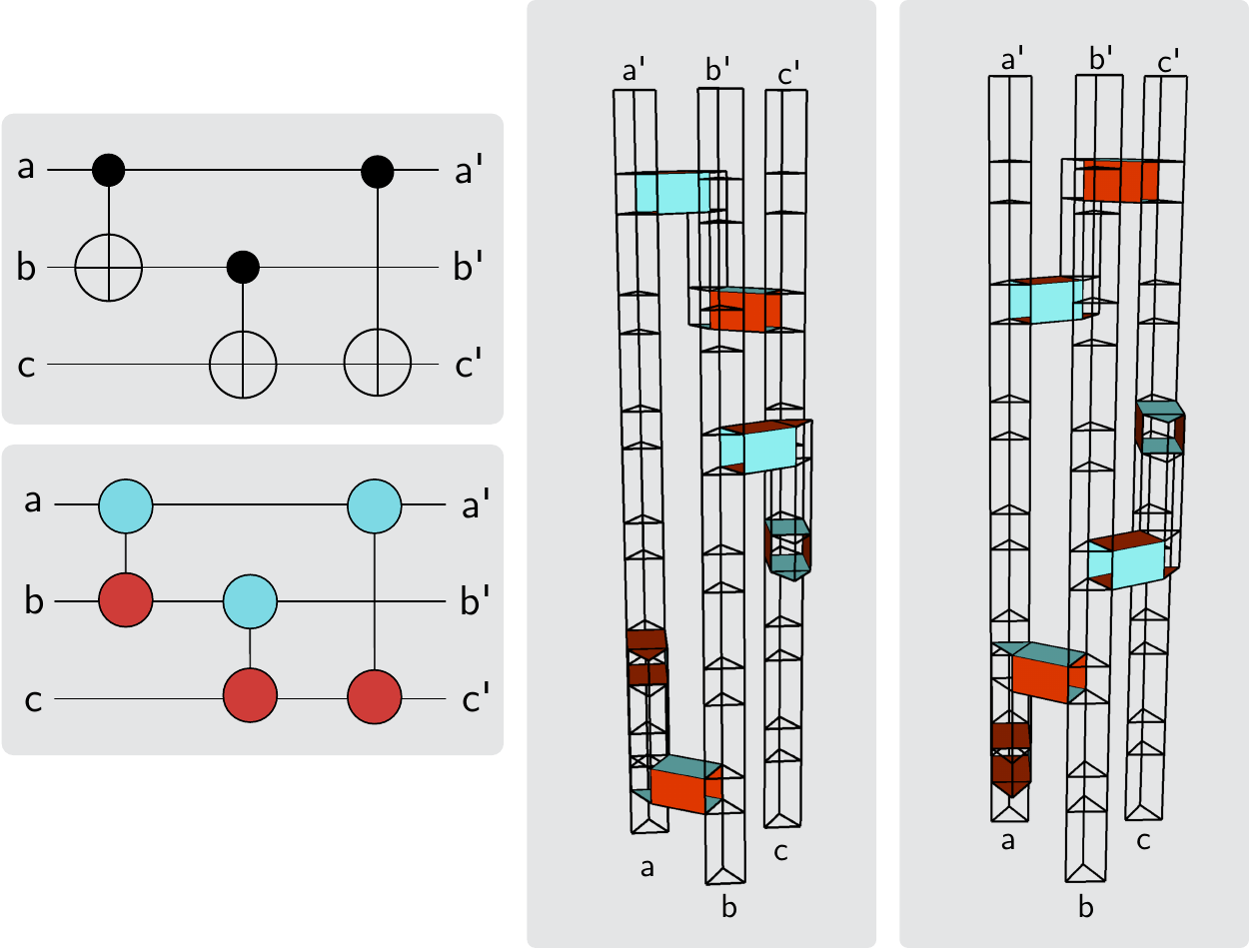}
    \caption{Three consecutive CNOT gates as a circuit and translated into a ZX diagram (left). Naively stacking the two variations of a single CNOT gate from \autoref{fig:pipes_cnot_nocs} onto each other, respectively (middle, right).}
    \label{fig:3cnot_naive}
\end{figure}
In order to understand the strength of the pipe diagrams, consider an example of three CNOT gates that are supposed to be applied as seen in \autoref{fig:3cnot_naive} (left), both as a circuit and as a ZX diagram. For the moment, let us assume that we aim for a representation with temporal open ports with usual time ordering first. Given the knowledge of the previous section, the construction of a CNOT as a pipe diagram, one can naively stack the CNOT gates as drawn in \autoref{fig:pipes_cnot_nocs} onto each other as shown in \autoref{fig:3cnot_naive} (middle, right). From the fact that we count only the effectively necessary prisms into the spacetime volume, it is already imminent that this representation is taking up far more volume than necessary. Ignoring this for the moment for illustrative purposes and considering the structures as drawn, they yield an effective $STV=18$ (middle) and $STV=16$ (right). %

\begin{figure}[t]
    \centering
    \includegraphics[width=\linewidth]{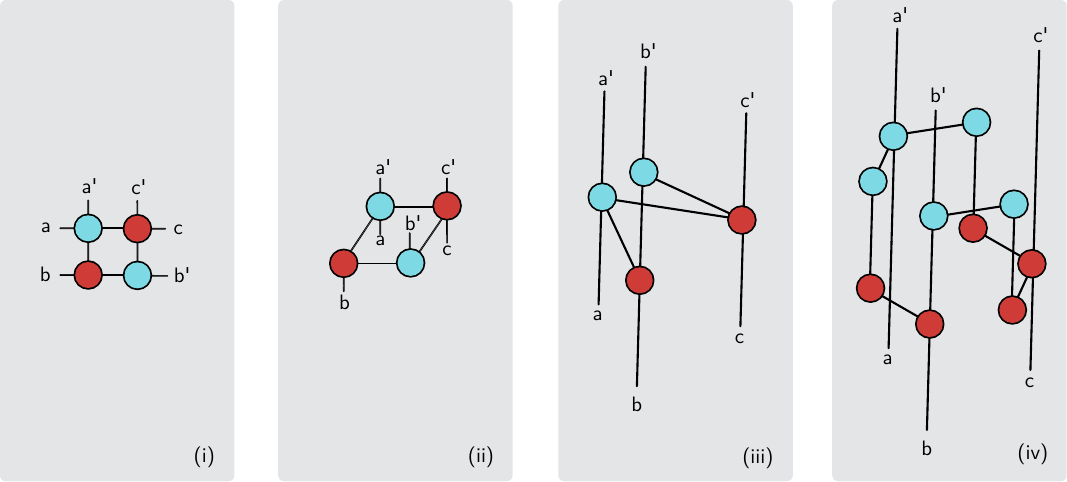}
    \caption{Having reduced the 3 CNOTs in terms of ZX diagrams via spider fusion (f) and rearranging the nodes, one has to determine how to embed this compact ZX diagram in spacetime.}
    \label{fig:embedding_3cnots}
\end{figure}

\begin{figure}[t]
    \centering
    \includegraphics[width=0.7\linewidth]{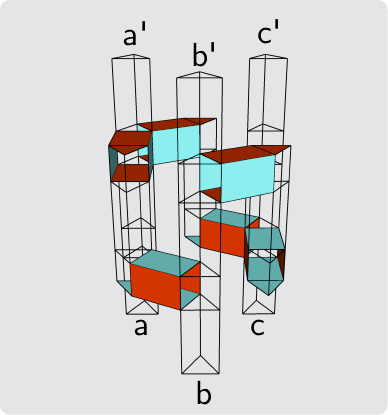}
    \caption{Given the ZX diagram from \autoref{fig:embedding_3cnots}, a direct translation into a pipe diagram is possible.}
    \label{fig:final_3cnots_pipediagram}
\end{figure}

As a next step, we reduce the required spacetime volume by exploiting the properties of the pipe diagram representation. While it is possible to directly transform the given pipe diagrams from \autoref{fig:3cnot_naive} (middle, right), the subsequent description follows the logic from \autoref{fig:overview} for illustrative purposes. Given the ZX diagram of the desired computation \autoref{fig:3cnot_naive} (left), one can apply spider fusion (f) two times and rearrange the nodes and wires to arrive at the representation in \autoref{fig:embedding_3cnots}{(i)}. This is a very compact representation of the computation; however, to embed it in spacetime (with open temporal ports) it is advisable to transform the ZX diagram with the available spacetime structure in mind. In the case of the color code, nodes can be placed on a hexagonal lattice in the spatial plane, with time flowing from bottom to top. Thus, \autoref{fig:embedding_3cnots}{(ii)} is a first step towards that direction in which we align the nodes along a spatial plane and the incoming and outgoing open wires are aligned along the temporal direction. Similarly to the case of the construction of the CNOT gate as a spacetime diagram (\autoref{subsec-cnot}), one needs to add identity nodes, i.e. auxiliary patches, to enable the translation of the structure into a pipe diagram. To enable this, \autoref{fig:embedding_3cnots}{(iii)} further positions the nodes in space, taking the positions of auxiliary nodes into account. Finally, \autoref{fig:embedding_3cnots}{(iv)} displays a ZX diagram with nodes aligned on a hexagonal lattice on the spatial plane, which can be directly translated into a pipe diagram. The respective pipe diagram is shown in \autoref{fig:final_3cnots_pipediagram}. 

The spacetime volume of this computation is much smaller than the volume of the naive, sequential application of CNOT gates. Either by counting the prisms that are not trivially ending in an open port in \autoref{fig:final_3cnots_pipediagram} or the nodes in \autoref{fig:embedding_3cnots}{(iv)}, one can conclude that the effective spacetime volume occupied by this structure is $STV=10$. Note that the same spacetime volume can be reached by using a pipe diagram for the surface code.

\begin{figure}
    \centering
    \includegraphics[width=0.8\linewidth]{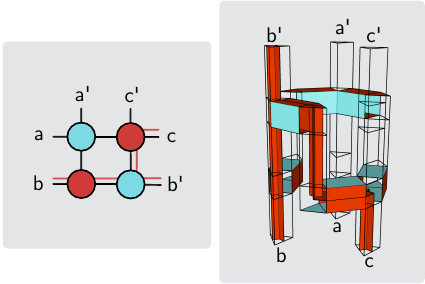}
    \caption{An example correlation surface of the compact ZX diagram with temporal ports only and its equivalent in the final pipe diagram.}
    \label{fig:compact_3cnot_example_CS}
\end{figure}

To sanity-check the pipe diagram's logic, one can explicitly draw the correlation surfaces and compare whether they coincide with the correlation surface given by the ZX diagram. An example correlation surface in both structures is shown in \autoref{fig:compact_3cnot_example_CS}.

\begin{figure}[t]
    \centering
    \includegraphics[width=\linewidth]{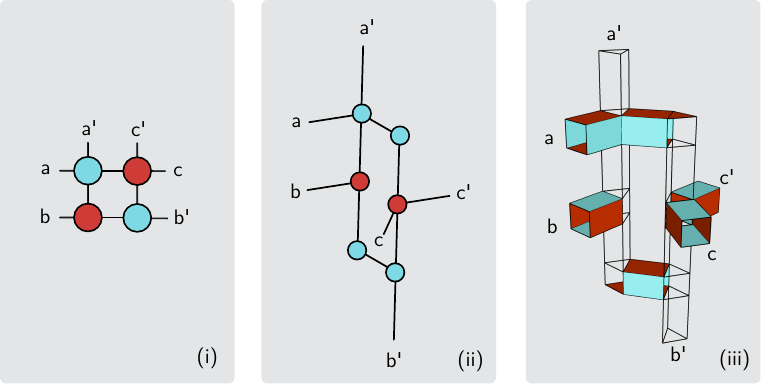}
    \caption{With the compact ZX diagram of the 3 CNOTs (i) one can embed the ZX diagram in space more directly if open ports in space and time are allowed (ii, iii).}
    \label{fig:3cnots_spatial_ports}
\end{figure}

\begin{figure}
    \centering
    \includegraphics[width=0.9\linewidth]{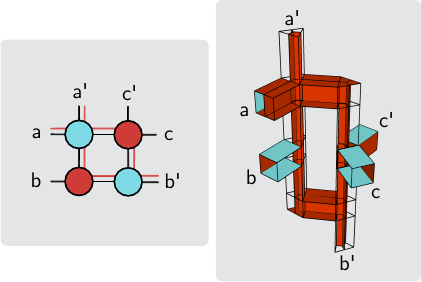}
    \caption{An example correlation of the compact ZX diagram with spatial and temporal open ports and its equivalent in the final pipe diagram. For better visibility some walls are removed.}
    \label{fig:3cnots_spatial_ports_CS}
\end{figure}

However, this construction is not minimal in spacetime volume. As the roles of space and time can be lifted to some extent in the picture of pipe diagrams, one can embed the compact ZX diagram in spacetime with lower $STV$ as soon as we allow not only temporal open ports but also spatial open ports. Starting from the same compact ZX diagram in \autoref{fig:3cnots_spatial_ports}{(i)}, one adds two identity nodes, as horizontal pipes can only connect nodes of the same color in the bulk~\footnote{Of course, if a spatial Hadamard was applied within a spatial pipe, blue and red prisms can be connected also in the bulk of a diagram. But this adds logical action which is not desired here.}, and places the open ports in space and time accordingly as shown in \autoref{fig:3cnots_spatial_ports}{(ii)}. This directly translates to the pipe diagram in \autoref{fig:3cnots_spatial_ports}{(iii)}. A sanity-check of an example correlation surface is shown in \autoref{fig:3cnots_spatial_ports_CS}. The spacetime volume of this computation is reduced down to $STV=6$. Also in the surface code, a pipe diagram with spatial and temporal open ports can be reduced down to $STV=6$. 

One should note that such reductions in spacetime volume are a degree of freedom that is often ignored in lattice surgery compilation literature~\cite{molavi_dependency-aware_2024, herzog_lattice_2025, herzog2025exploitingmovablelogicalqubits} that assumes a fixed lattice surgery implementation of the CNOT gate and thus just naively stacks the gates as shown in \autoref{fig:3cnot_naive}. In terms of pipe diagrams, reductions of the spacetime volume are apparent and arise naturally.

\subsection{Single-Qubit Cliffords}\label{subsec-compare-sc-cc}
\begin{figure}
    \centering
    \includegraphics[width=\linewidth]{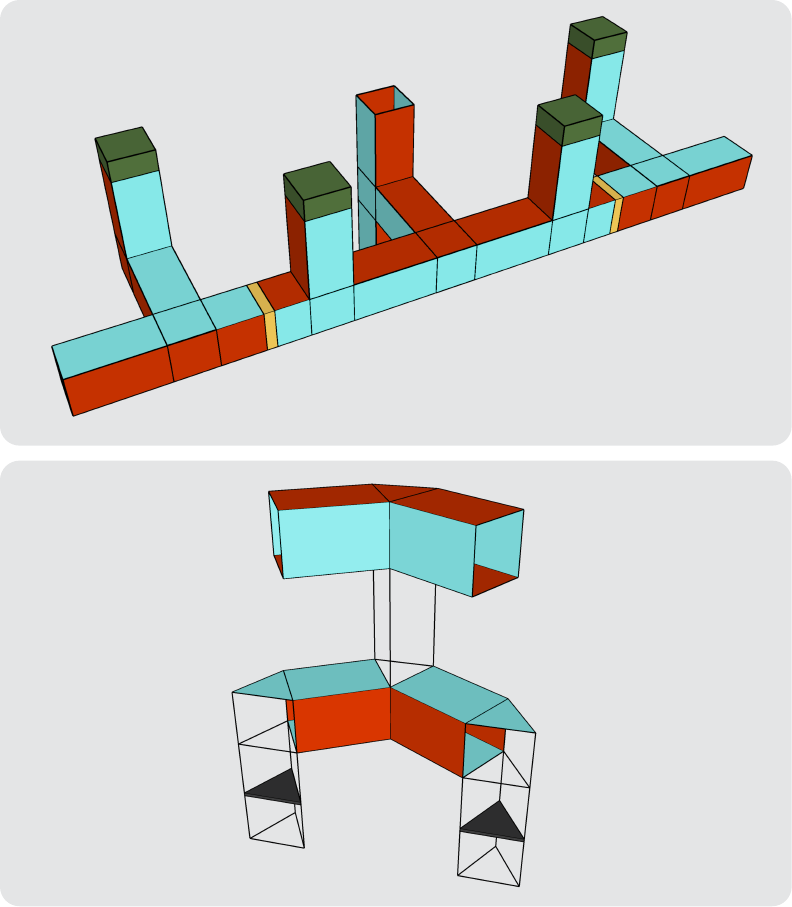}
    \caption{CNOT gate together with single-qubit Clifford gates on the target qubit for the surface code (top) and the color code (bottom). The diagrams amount to the unitary $U =S(1)H(1)S(1)CNOT(0,1)S(1)H(1)S(1)$. %
    }
    \label{fig:sc_cc_clifford}
\end{figure}

As pointed out previously in \autoref{subsec-full-clifford}, color codes have the decisive benefit that they admit a transversal application of logical single-qubit Clifford gates, generated by $S$ and $H$ gates. To illustrate the benefit in an explicit example, consider the logical quantum circuit defined by the unitary $U =S(1)H(1)S(1)CNOT(0,1)S(1)H(1)S(1)$. 

A low-overhead representation for the surface code, as shown in \autoref{fig:sc_cc_clifford} (top), can be found by using the version of the CNOT gate that has two open spatial and two open temporal pipes. One applies the $S$ and $H$ gates along the spatial direction via $Y$-basis measurements as introduced in Fig. 9 of \cite{Gidney2024inplaceaccessto}, indicated by a green half-cube each. The Hadamard gates can also be applied in the spatial direction indicated by a yellow band within the connecting pipes. Such a spatial Hadamard can be performed by using mixed $Z$- and $X$-type stabilizers at the interface of two patches. Note that the $Y$ basis measurement for performing an $S$ gate is not agnostic about the wall colors of the involved pipes and cubes, which is the reason why the implementation on the left and right sides of the spatial Hadamards has to be different, respectively. One has to ensure that the cube in which the $S$ gate ``enters'' has to be equivalent to a blue ZX node. To fulfill this requirement, the spacetime overheads for $S$ gates at different points in the diagram differ. Overall, this diagram amounts to a spacetime volume of $STV = 8 + 4\cdot\frac{1}{2}=10$ by counting the cubes and half-cubes.

If this gate sequence is embedded in a larger quantum circuit, it may be necessary to rotate the altered boundaries back to the original orientation, which can be done via patch deformation (e.g. by forming an L-bend). On the other hand, the induced flip of wall colors may also help to find a more efficient representation locally.

For the color code, a low-overhead construction of $U$ can be done by using the compact CNOT gate with open spatial ports and by adding vertical pipes with a black plane, which represent multiple transversal gate applications as shown in the bottom of \autoref{fig:sc_cc_clifford}. %
By counting the prisms included in this diagram, the spacetime volume is $STV = 4$~\footnote{Note that the application of spatial gates would possibly further reduce the spacetime volume.}. 

Overall, this small example displays that there is potential to reduce overhead for single-qubit Clifford gates by using the color code. %

\subsection{Summary of Benefits and Deficits of the Color Code}~\label{subsec-benefits-cc}
This summarizes how pipe diagrams can be created for the color code setup with triangular patches on the 6.6.6 code. At this point let us summarize the overall benefits and deficits of the color code in comparison to the surface code, beyond the previously discussed benefit of transversal $S$ and $H$ gates.

For a $d=2t+1$ color code patch, there are \mbox{$q = 3t^2 + 3t+1$} data qubits involved. In comparison, for the rotated surface code $q = 4t^2 + 4t +1$. In particular for large-scale computations any difference in the number of qubits will be relevant for the practicality of a quantum error correcting code. Ultimately, the qubit overhead needed to reach a given logical error rate is what matters -- and since color code decoding is harder, it is still an open question in decoding research which scheme is effectively more efficient regarding the qubit count.%

Furthermore, in larger scale computations the fact that higher-rank junctions are possible in the color code in contrast to the surface code (\autoref{subsec-zx-junctions}) opens interesting directions for future investigation. A related aspect is the fact that the color code allows to access both $Z$ and $X$ logical operators from each side of the triangle, while the surface code allows access only to one type of logical operator. Thus, patch rotations are never necessary for the color code. Nevertheless, one should also mention that the color code patches are placed on a 3-valent connectivity graph in space, while the surface code patches can be placed on a 4-valent connectivity graph in space. The spatiotemporal connectivity of a given pipe diagram is, however, described by a 5-valent graph for the color code and a 4-valent graph for the surface code. Furthermore, note that in the color code only two nodes of the same color can be directly connected horizontally in the bulk of a diagram, which may add spacetime overhead. The interplay of these aspects for larger scale computations is left for future studies.

Even though the current discussion is restricted to Clifford operations only, we note that magic state cultivation~\cite{gidney_magic_2024} is considerably easier for the color code than for the surface code, as one can skip the grafting procedure. 

The major reason why the surface code remains the predominant code in quantum error correction research is: The decoding problem is much easier and quicker to solve for the surface code due to its matchability. Color codes lack the property of matchability, which makes them much harder and, in particular, slower to decode. However, there is progress towards making color code decoding competitive in the context of real-time decoding~\cite{koutsioumpas2026_gpu}.

\section{Microscopic Compilation: Syndrome Extraction for Color Codes}\label{sec-microscopic-compilation}
The choice of syndrome extraction scheme can significantly impact the performance of a quantum error correcting code, as the physical implementation introduces error mechanisms that are not visible at the stabilizer level of abstraction. In particular, so-called hook errors -- errors arising from the particular order of physical CNOT gates during stabilizer measurements -- pose a serious challenge and must be carefully addressed in the syndrome extraction circuit.

\subsection{Syndrome Extraction}

In the surface code, the geometric structure of the stabilizers and logical operators allows to reduce the impact of hook errors by reordering the CNOT gates during syndrome extraction in such a way that hook errors propagate perpendicular to the respective logical operator -- such that they cannot do harm, as for instance explained in~\cite{yoder_surface_2017}. Alternatively, using a different schedule, hook errors can be oriented along diagonals of the surface code plaquettes such that they never align with logical operators which always traverse horizontally or vertically~\cite{kishony2026surfacecodeoffthehookdiagonal}. As explained before, the color code does not have the structure of perpendicular logical operators, such that a similar strategy cannot be applied here. Instead, other strategies are applied, such as incorporating flag qubits~\cite{Chamberland_2020_flag_qubits}. However, more compact circuits such as the superdense syndrome extraction scheme and the inline folding scheme~\cite{gidney2023new} can be beneficial in comparison. It has also been shown recently that all malign hook errors in the bulk of the color code can be avoided by using different gate schedules for plaquettes of different colors~\cite{kishony2026colorcodeoffthehookavoiding}.

\begin{figure}[t]
    \centering
    \includegraphics[width=\linewidth]{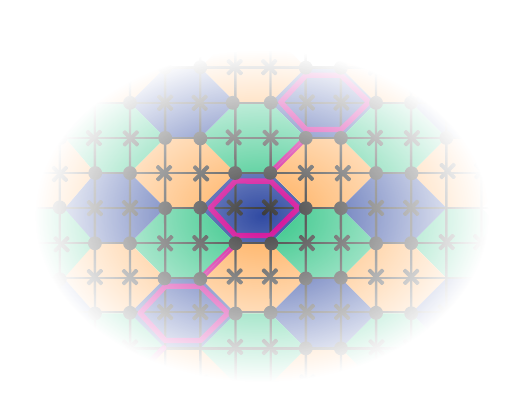}
    \caption{Physical data and auxiliary qubits are assumed to have nearest-neighbor connectivity on a square lattice with a few diagonal couplings. Data qubits are marked with a black circle and auxiliary qubits with black crosses. The plaquettes or lines in pink depict single-type stabilizers along an STDW.}
    \label{fig:se_geometry_cc}
\end{figure}

\begin{figure}
    \centering
    \includegraphics[width=0.5\linewidth]{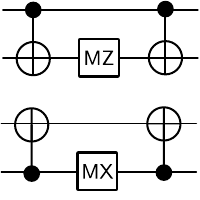}
    \caption{Folding syndrome extraction purely on data qubits for weight-2 stabilizers without usage of auxiliary qubits for $Z$ stabilizer measurement (top) and $X$ stabilizer measurement (bottom).}
    \label{fig:se_weight_two}
\end{figure}

The superdense and folding schemes are adopted in the current setup. Similar to~\cite{gidney2023new}, we place the physical data qubits on a square lattice with nearest-neighbor connectivity as shown in \autoref{fig:se_geometry_cc}. Additionally, we assume diagonal nearest-neighbor connectivity between selected qubits at the lattice. The figure shows a cutout of arbitrarily large color code patches at the location of an STDW: Pink outlines depict single-type stabilizers of weight-6 or weight-2, where black circles at the vertices of the grid depict physical data qubits and black crosses are physical auxiliary qubits. For the double-type stabilizers, it is straightforward to adopt the superdense syndrome extraction scheme~\cite{gidney2023new}, which requires two auxiliary qubits per patch. In this scheme, the two auxiliary qubits are entangled in a Bell state such that CNOT gates between the auxiliary and data qubits can be applied such that the nearest-neighbor restrictions remain satisfied. Both the $X$ and $Z$ stabilizers can be measured with entangling and disentangling the auxiliary qubits once. With distance $d$ on the stabilizer level, also the circuit level distance is $\Tilde{d}=d$. %
The single-type stabilizers along the STDW of weight-3, -5 and -6 can be handled with the same scheme, but with shorter schedules, as only one stabilizer is measured instead of two. The situation changes fundamentally when the single-type weight-2 stabilizers are considered: While all other stabilizers have a pair of physical auxiliary qubits that are assigned to them, this is not the case for the single-type weight-2 stabilizers. Using auxiliary qubits that are already utilized by other stabilizers either requires lifting the nearest-neighbor constraint (if the auxiliary qubits of single-type, higher weight stabilizers are supposed to be used) or making the schedule deeper (if auxiliary qubits of double-type stabilizers are supposed to be used), which scales with the rounds of syndrome extraction. As none of the possibilities is desirable, we choose to use the inline folding~\cite{gidney2023new} scheme for the single-type weight-2 stabilizers, which does not require any auxiliary qubits and works solely on the data qubits. The explicit circuits for this scheme in the present case are shown in \autoref{fig:se_weight_two}. Even though this scheme can reduce the circuit-level distance to $\Tilde{d}={\lceil d/2\rceil}$, we choose this type of syndrome extraction as compactness of the circuits plays a crucial role for the performance as well~\cite{gidney2023new}. Furthermore, the distance is only reduced in the two-dimensional regions occupied by the STDWs, such that the effect on the logical error rate is expected to diminish at large distances.

\begin{figure*}[t]
    \centering
    \includegraphics[width=0.9\linewidth]{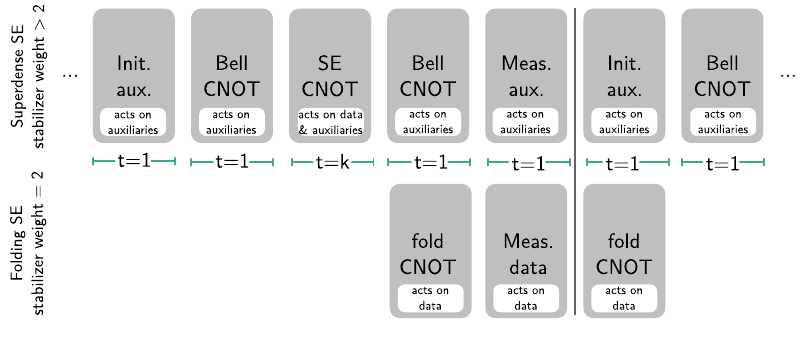}
    \caption{How the superdense and folding syndrome extraction scheme are interleaved such that only a constant time overhead is generated. Abbreviations are explained in the main text.}
    \label{fig:interleave_se}
\end{figure*}

Combining both the superdense and inline folding scheme can be done with maximally a constant (distance-independent) time overhead of a single time step~\footnote{One timestep is defined by a multitude of gates that can be applied in parallel, i.e., with disjoint physical qubit overlap.}. For a schematic on how both schemes can be interleaved, consider \autoref{fig:interleave_se}. The top row schematically shows the steps during superdense syndrome extraction: One starts with the initialization of the auxiliary qubits in the $\ket{+}$ and $\ket{0}$ states respectively (``Init. aux.''), which is followed by the entangling CNOT between the auxiliary qubits (``Bell CNOT''). Then, the CNOT gates between data and auxiliary qubits (``SE CNOT'') are applied. The number of timesteps in that step is $t=k$ with $k=6$ for a weight-6 double-type stabilizer. This is followed by the disentangling CNOT gate on the auxiliaries (``Bell CNOT'') and is concluded with a measurement in $X$ and $Z$ bases respectively (``Meas. aux.''), which yields the syndrome measurement's outcome for the $X$ and $Z$ stabilizers. After the black vertical line, a new round of syndrome extraction begins. The bottom row of \autoref{fig:interleave_se} shows how the three steps of the inline folding for weight-2 single-type stabilizers can be interleaved, where ``fold CNOT'',  ``Meas. data'' and again ``fold CNOT'' refers to the three steps as shown in \autoref{fig:se_weight_two}. As the gates in these steps act on the data qubits, the steps must be performed simultaneously with steps of the superdense scheme that act on the auxiliary qubits only to avoid conflicts. The earliest possible way is shown in the figure -- which overlaps with the next round of syndrome extraction by a single timestep. Thus, the combination of schemes can maximally lead to a single timestep overhead (independent of $d$) if the single-type stabilizers have to be measured at the end of the overall computation. Furthermore, note that for the double-type stabilizers during the superdense scheme, one has to apply $X$ gates based on classical feedback, which is not taken into account in the present considerations, as we assume that Pauli gates are not executed physically but tracked in software.

Finally, note that syndrome extraction using only the connectivity of the square lattice is certainly possible, though potentially at the cost of increased circuit depth.

\subsection{Simulations}

\begin{figure*}[ht]
    \centering
    \subfloat[]{\includegraphics[width=0.48\linewidth]{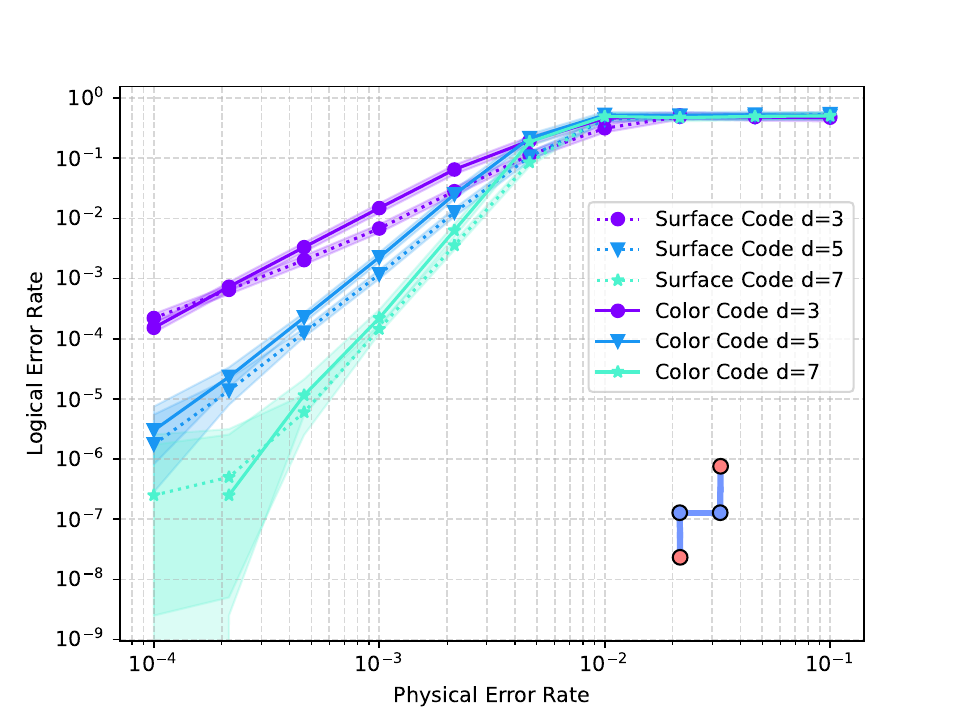}
        \label{fig:simulation_a_tele}}
    \hfill
    \subfloat[]{\includegraphics[width=0.48\linewidth]{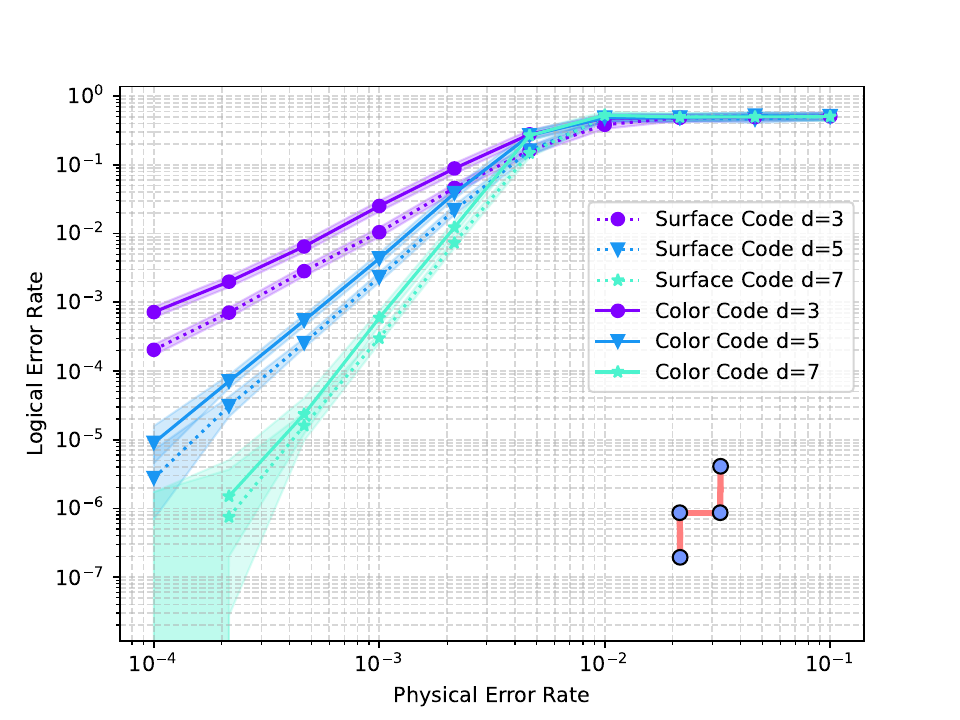}
        \label{fig:simulation_b_tele}}
    \hfill
    \subfloat[]{\includegraphics[width=0.48\linewidth]{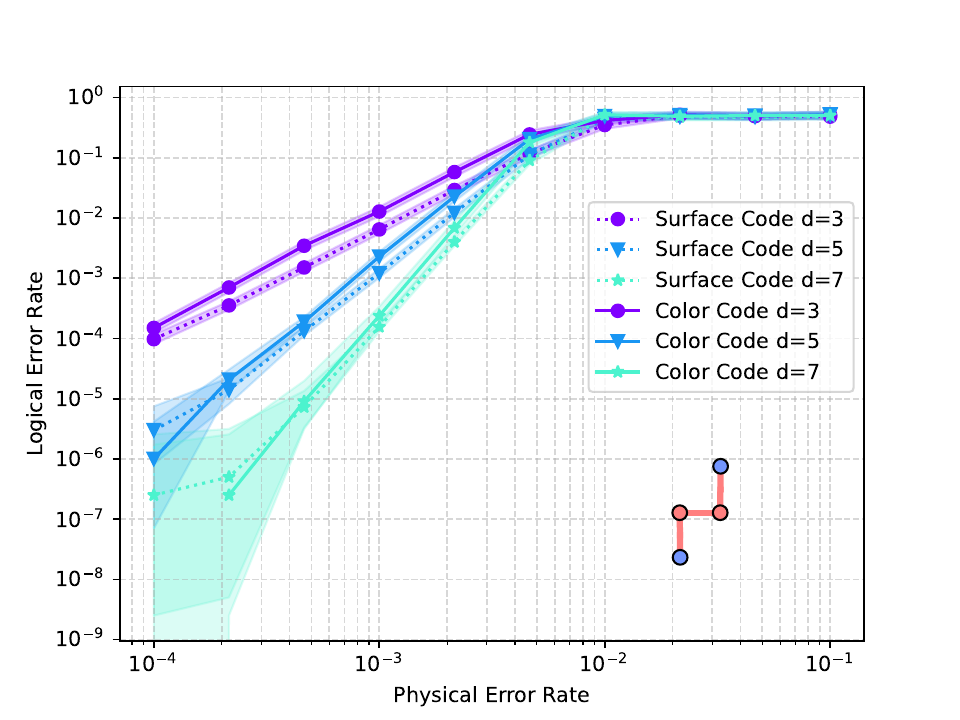}
        \label{fig:simulation_c_tele}}
    \hfill
    \subfloat[]{\includegraphics[width=0.48\linewidth]{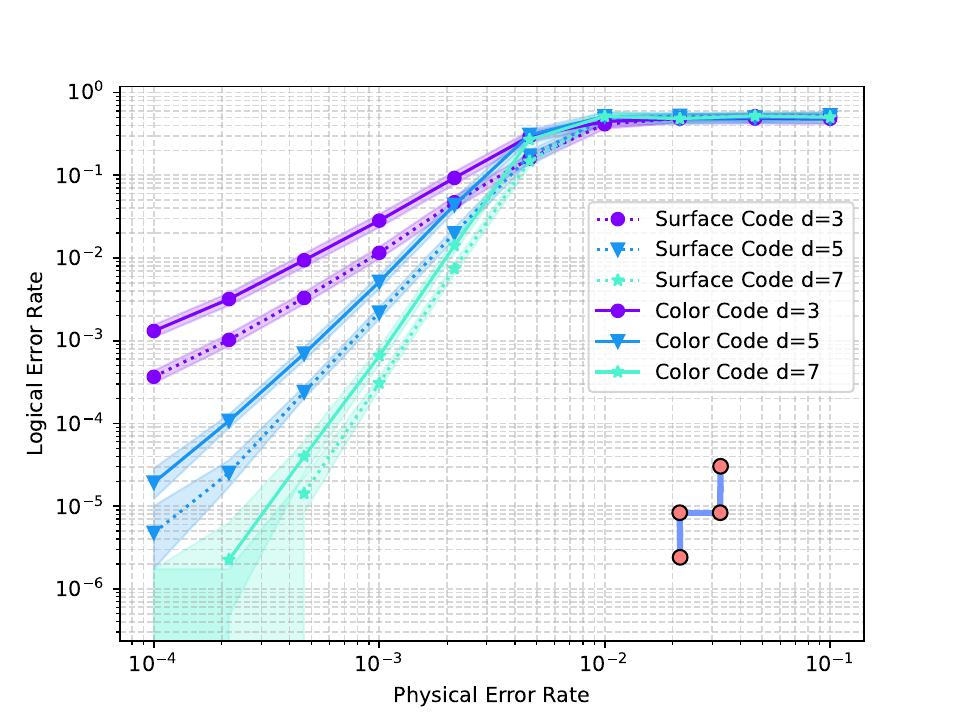}
        \label{fig:simulation_d_tele}}
    \caption{Comparison of logical error rates between the surface code and color code for a movement for all four possibilities of movement setups of this kind. For the color code, the tesseract decoder was used, and for the surface code, correlated pymatching. The pipe diagram from \autoref{fig:teleportation} represents the cases (c) and (d) for different initialization and measurement, respectively.}
    \label{fig:simulation_a_tele_sc_cc}
\end{figure*}

\begin{figure*}[ht]
    \centering
    \subfloat[]{\includegraphics[width=0.48\linewidth]{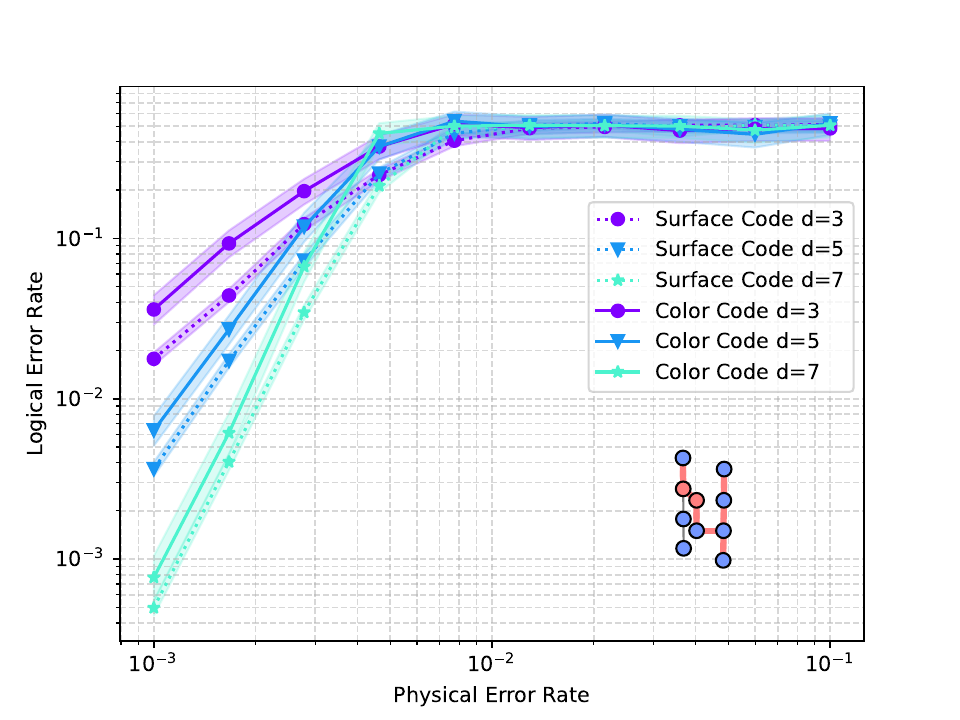}
        \label{fig:simulation_ab_obs0_cnot}}
    \hfill
    \subfloat[]{\includegraphics[width=0.48\linewidth]{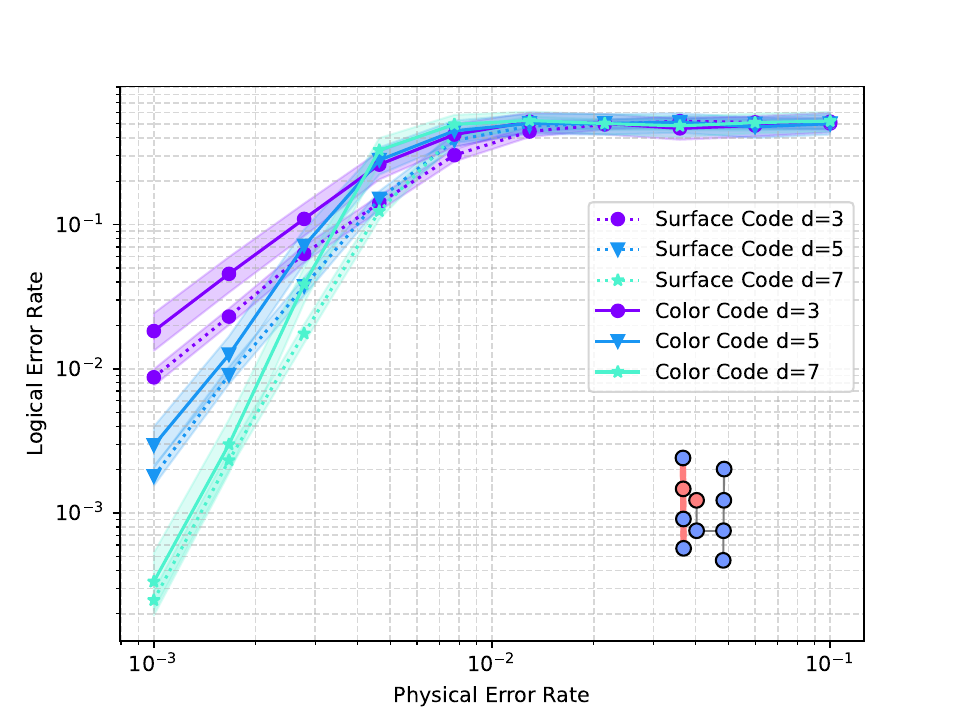}
        \label{fig:simulation_ab_obs1_cnot}}
    \hfill
    \subfloat[]{\includegraphics[width=0.48\linewidth]{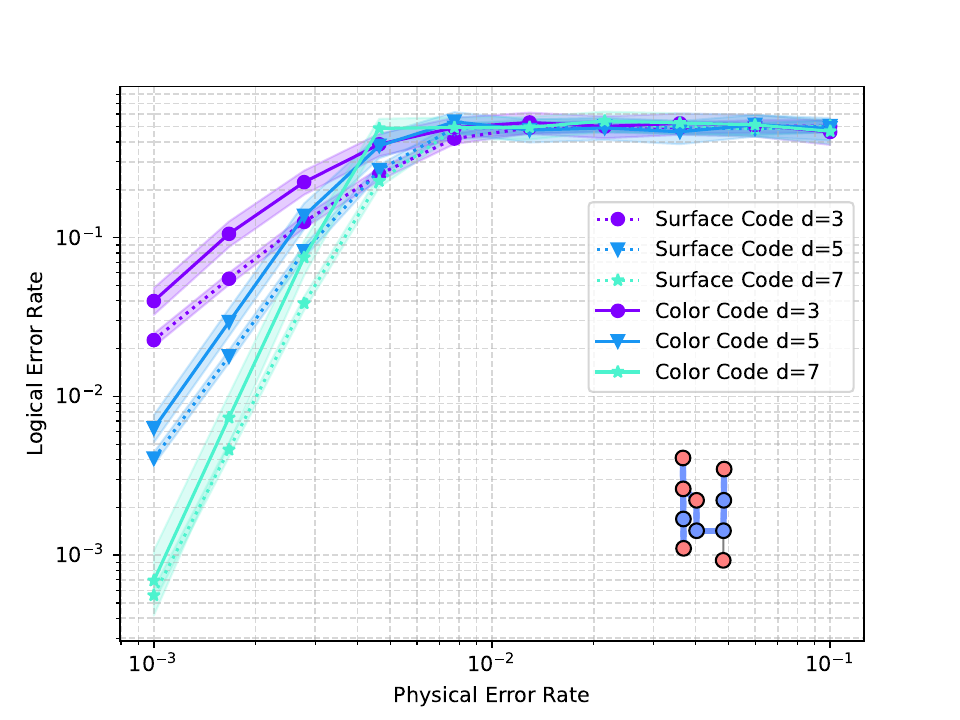}
        \label{fig:simulation_cd_obs1_cnot}}
    \hfill
    \subfloat[]{\includegraphics[width=0.48\linewidth]{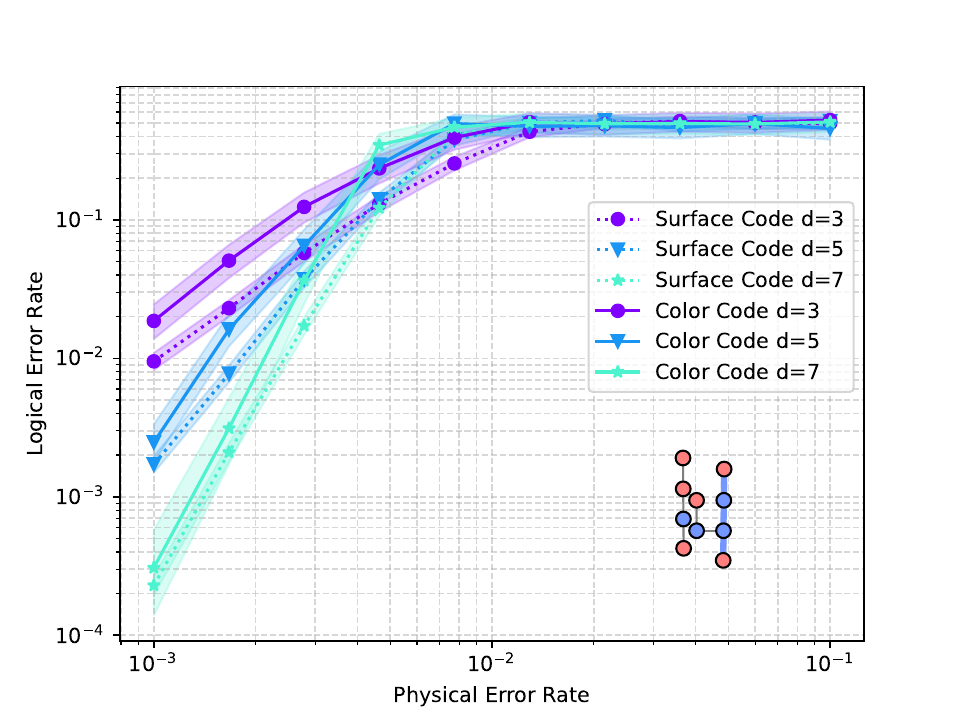}
        \label{fig:simulation_cd_obs0_cnot}}
    \caption{Comparison of logical error rates between the surface code and color code for a logical CNOT gate with the different available logical operators. For the color code, the tesseract decoder was used, and for the surface code, correlated pymatching. The pipe diagram in \autoref{fig:pipes_cnot_nocs} (right) displays what is simulated here.}
    \label{fig:cnotsim}
\end{figure*}

We simulate the aforementioned syndrome extraction scheme for chosen examples of pipe diagrams with \texttt{stim}~\cite{gidney2021stim}. The respective circuits for the equivalent pipe diagrams in the surface code are generated with \texttt{tqec}~\cite{tqec} and decoded with correlated \texttt{pymatching}~\cite{Higgott2025sparseblossom, fowler2013optimalcomplexitycorrectioncorrelated}. The syndrome extraction circuits for the color code are decoded with the \texttt{tesseract} decoder~\cite{beni2025tesseractdecoder}. Throughout the simulations, we employ the uniform circuit-level depolarizing noise model, i.e., with the same probability $p$, idling qubits depolarize, after each 1- and 2-qubit gate there is 1- and 2-qubit depolarization, as well as flipped measurement results and reset errors. Simulation results and code to reproduce the results can be found at \url{https://github.com/LSHerzog/tqec}.

Simulation data for the movement of a single patch, similar to \autoref{fig:teleportation}, is shown in \autoref{fig:simulation_a_tele_sc_cc}. There are two possibilities to move a patch by one position since there are two possible alignments of the colors on the horizontal pipe as well as two different initialization and measurement possibilities. For both options, one can either simulate the behaviour of the $X_L$ or $Z_L$ operator, such that there are four options in total, which are shown in the figure. Shaded regions show hypotheses with likelihoods within a factor 1000 of the maximum likelihood hypothesis, given the sampled data~\footnote{An explanation of binomial confidence intervals in stim can be found \href{https://tqec.github.io/tqec/user_guide/reading_error_plots.html\#confidence-intervals-and-missing-points}{here}.}.

As expected, the surface code performs better overall. However, these simulations are not intended to make a statement about the superiority or inferiority of either code — decoder performance plays a significant role in this comparison, and optimizing decoders is beyond the scope of this work. Rather, the results are meant to serve as a proof-of-principle, illustrating that logical computation with the color code remains a viable and worthwhile endeavor.

One can observe that \autoref{fig:simulation_b_tele} and \autoref{fig:simulation_d_tele}, in particular for $d=3$ and $d=5$, the difference between the surface code's and color code's performance differs significantly more than in the case of \autoref{fig:simulation_a_tele} and \autoref{fig:simulation_c_tele}. For this, it is vital to know which type of correlation surface is taken into account: The first pair corresponds to a pipe diagram with a horizontal correlation surface in the horizontal pipe, while the latter pair corresponds to a vertical correlation surface in the horizontal pipe. As mentioned in the previous section already, the folding scheme for the weight-2 stabilizers decreases the overall circuit-level distance and creates problematic spatial errors. Thus, these errors propagate perpendicularly in comparison to the vertical correlation surface and can therefore induce flips in the correlation surface's measurement results. Thus, the circuit-level distance is reduced $\Tilde{d} = \lceil d/2 \rceil$. This mechanism does not apply if a horizontal correlation surface is taken into account, where the circuit level distance $\Tilde{d} = d$ with the distance $d$ of the separate patches. This mechanism may be the reason for the higher difference for $d=3$ and $d=5$ if a vertical correlation surface is present. However, this primarily applies to low distances and is expected to diminish for high distances, as this effect happens only along the STDW but not within the whole bulk of the color code patches. 

A CNOT gate with temporal open ports is simulated in \autoref{fig:cnotsim}. One can observe that \autoref{fig:simulation_ab_obs1_cnot} and \autoref{fig:simulation_cd_obs0_cnot} for both the surface and color code have the tendency to achieve lower error rates than \autoref{fig:simulation_ab_obs0_cnot} and \autoref{fig:simulation_cd_obs1_cnot} due to the fact that in the first pair, the correlation surface is smaller and hence errors from correlation surfaces in horizontal pipes are not entering. %
Furthermore, there is no significant difference between \autoref{fig:simulation_ab_obs0_cnot} and \autoref{fig:simulation_cd_obs1_cnot} as both contain both a vertical and a horizontal correlation surface in the horizontal pipes each. Thus, the effects of the vertical correlation surface in the horizontal pipe cannot be differentiated.

\section{Conclusion}\label{sec-conclusion}
In this work, we developed a pipe diagram representation for the triangular color code on the 6.6.6 lattice based on a lattice surgery scheme with semi-transparent domain walls, valid for arbitrary code distance $d$. From this microscopic construction, distance-independent pipe diagrams emerge alongside constructions of correlation surfaces, and we showed how these connect to a ZX-diagrammatic description of the computation. Building on these foundations, we presented examples of macroscopic compilation that illustrate the potential of this approach, in particular with the availability of transversal single-qubit Clifford gates. Also the higher-degree ZX nodes in the color code may be a benefit in this picture. At this stage, however, it is not yet possible to make a general statement about whether the color code or the surface code performs better in terms of spacetime overhead; this work is intended as a proof-of-principle exploration of pipe diagrams for a code beyond the surface code. We further proposed a syndrome extraction scheme for this setup and showed via simulation that the resulting logical error rates are reasonable in comparison to the surface code, though faster and more accurate decoders will be necessary to make this comparison more conclusive.

Future work should explore pipe diagram constructions based on other lattice surgery schemes, such as those of~\cite{thomsen_low-overhead_2024}, which are not considered here. Moreover, exploring topological codes with multiple logical qubits encoded per patch or color codes on lattices other than the 6.6.6 lattices would be worthwhile to explore too. Furthermore, alternative syndrome extraction schemes should be explored and quantitatively compared. A further important direction is the development of fully automated software for both microscopic compilation, including syndrome extraction circuit generation, and macroscopic compilation, i.e., the compact embedding of ZX diagrams as pipe diagrams in spacetime~\cite{topologiq, zhou2026topolslatticesurgerycompilation}. Finally, this work is restricted to Clifford operations; incorporating $T$ gates into the color code pipe diagram framework will be an important next step towards universal fault-tolerant computation.

\section*{Acknowledgments}
L.H. thanks Lucas Berent, Aleksander Kubica, Tom Peham, Ludwig Schmid, and Erik Weilandt for helpful discussions and feedback on the project. 3D figures were created with SketchUp. We used \href{https://piper-draw.walruscomputing.com/}{Piper Draw} by Walrus Computing to check surface code diagrams in the process of this work. L.H. and R.W. acknowledge funding from the European Research Council (ERC) under the European Union’s Horizon 2020 research and innovation program (grant agreement No. 101001318) and Millenion (grant agreement No. 101114305). This work was part of the Munich Quantum Valley, which is supported by the Bavarian state government with funds from the Hightech Agenda Bayern Plus. This work was funded by the Deutsche Forschungsgemeinschaft (DFG, German Research Foundation, No. 563402549). Furthermore, this work was supported by the BMFTR under grant number 13N17298 (SYNQ).

\bibliography{bibliography}

\end{document}